\documentclass{cshonours}
\usepackage{pgfplots}
\pgfplotsset{width=7cm,compat=1.8}
\usepackage{pgfplotstable}

\usepackage{sfmath}
\usepackage{graphics}  
\usepackage{enumitem}
\usepackage{amsmath}
\usepackage{mathtools}
\usepackage[]{algorithm2e}
\usepackage{bchart}
\usepackage{tikz,pgfplots}
\usepackage{subcaption}
\usepackage{graphicx}
\usepackage{sidecap}
\usepackage{color, colortbl}
\usepackage{multirow}
\usepackage{wrapfig}
\usepackage{hyperref}
\usepackage{xcolor}
\title{Scheduling Applications on Containers Based on  Dependency of The Applications}
\author{Abdullah Alelyani}
\begin{document}
\maketitle
\section*{Acknowledgement}
Nobody has been more supportive to me in during this time of this project than	the	members	of	my	family.	I would	like to thank my parents, who pray for me, and motivate me and they are with me in	whatever I pursue. Also, I would like to thank my loving	and supportive wife, Fatimah, and my three wonderful Kids, Elyass, Mohammed	and	Asya, who had been patient through the hard time that we are passing through since I left them in our country. I would like to say to them I miss you and I hope you will be able to come to Australia soon. Also, I would like to express my sincere thanks to my supervisors, Prof Amitava and Dr Mubashar who have been guiding, encouraging and supporting me throughout this project. I would stress my thanks to Prof Amitava for his patience and advice.
\abstract{
Cloud computing technology has been one of the most critical developments in provisioning both hardware and software infrastructure in recent years. Container technology  is a new  cloud technology that boosts the booting of applications, increases the ability to deploy applications on containers and improves the host machine resource sharing. Thus, enhancing a cloud container system needs a robust algorithm that deploys the applications efficiently. Most of the schedulers associated with container technology are focused on load balancing for increasing container performance. The traffic over networks plays a significant role in the performance of containers. Container deployment considering only load balancing may not be the best scheduling strategy due to the dependency between the applications that might be deployed in different pods (zones) in the container's cloud. This project aims to develop an algorithm that deploys applications into containers by considering the dependencies between applications as well as load balancing. The proposed algorithm performs better in terms of improving the throughput and reducing the network traffic as compared to state-of-the-art container scheduling algorithms.
}
\tableofcontents
\chapter{Introduction}
A container is a virtual space on a cloud machine's resources that runs applications. Each cloud machine runs a different number of containers depending on its resources. Container technology allows deploying applications on a cloud host machine and controlling the host resources to be shared between deployed applications. Managing application deployment on resources is the most important part of the container technology since appropriate deployment enhances the performance of the containers.  

Container technology uses a scheduler for deploying the applications whenever users want them to be run. Therefore, to boost the performance the scheduler should balance the loads between the containers. The optimum schedule is one that considers both the requirements of applications and the available resources. Moreover, the scheduler should consider other constraints such as response time, security, and overhead costs.  

Network traffic overhead influences the performance of containers. The network traffic overhead could be caused due to the network contentions \cite{8259462}. The network contention could be between any application that depends on other applications or services in the same zone or different zones. However, when dependent applications are distributed in different zones, the contentions, and the traffic overhead increase. Therefore, the deployment schedules of applications are responsible for reducing traffic overhead by considering the dependencies between the applications during the deployment stage.
Figure 1.1 demonstrates the network traffic between zones when the applications are distributed without considering the dependencies that connect them in a network. From the figure, it is assumed that application (A1), which is allocated to container 1, requests data from an application (A6), which is assigned to container 6 in the third host machine. If each request's cost is R and there are T requests, the total cost will be $RT$, and that is just for one application pair in the system.  Also, the responses for the requests also cost the same as the requests. The cost increases further when the application within a cloud container has more than one dependency. 
\begin{figure}[h!]
    \begin{center}
    \centering
    \scalebox{0.35}{
    \includegraphics{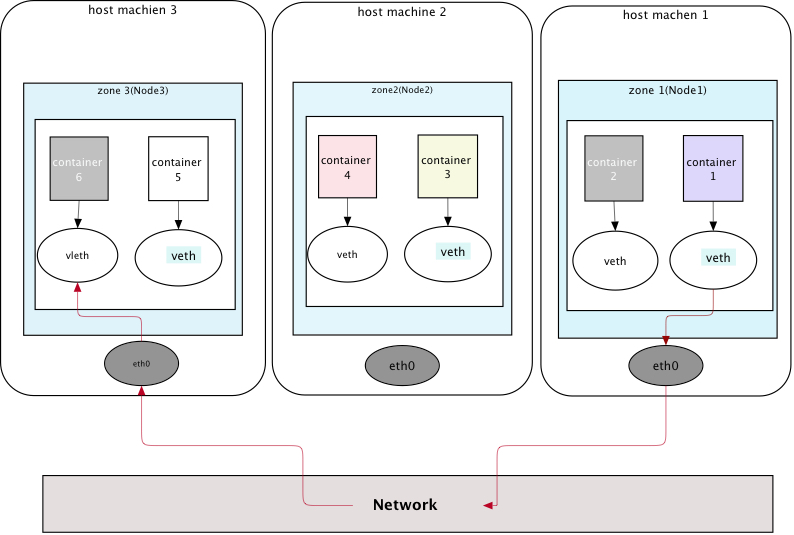}}
    \caption{Containers Communication Thought Network}
    \end{center}
\end{figure}
This research contributes to improving container performance by reducing the network traffic and maintaining the load balancing.
However, the cost of moving an application from a container to another container will be considered as a one-time cost, and this cost will be calculated based on the size of the application. 
The rest of this dissertation is organized as follows. Chapter 2 gives a brief background of container technology and literature review. In chapter 3, the problem and the research objectives will be defined. Chapter 4 will discuss the methodology of this research. Chapter 5 will discuss the experiments used for evaluating the  algorithm. Chapter 6 will introduce the results and analysis and a conclusion is provided in chapter 7.

\chapter{Literature Review}
\section{Motivation}
Cloud is a type of service that provides infrastructure and applications that can be utilized anytime, and from any place by users \cite{xiang2018method}. Moreover,  cloud services could be reached over the internet, and the services are delivered on demand \cite{duan2011modeling}.  The internet has expanded  utilization of cloud resources. Cloud technology’s goal is to meet user requirements of massive resources such as CPUs, RAMs, and hard disk spaces.
Therefore, virtualization technology was introduced to satisfy resource requests in the cloud. 
Virtualization is a technology that allows more efficient use of resources such as CPU and RAM by sharing them between applications. Many companies provide this technology, for example, Hyper-V from Microsoft, KVM from Red Hat, and VMware. Virtualization increases the utilization of cloud infrastructure and software. Furthermore, the flexibility and diversity that are provided by virtualization improve cloud utilization. Also, developing and managing network services and infrastructure increases the performance of the network that reflects on cloud performance\cite{duan2011modeling}. 
The advantage of the virtualization technology, along with managing the massive requests of users is in the fast access of a massive amount of resources such as data from the data centers \cite{younge2014advanced}. In addition, it allows users to reach parallel resources by using cloud interfaces.  According to Ananth \& Sharma \cite{ananth2017cost}, one of the most important benefits of virtualization is that it has been the best alternative solution to traditional network management. Finally, the service cost reduction that is made by virtualization increases the usage of  cloud computing systems.

\section{Introduction}
Although virtualization manages the multi-virtual machine by a hypervisor layer, it isolates the applications and has limited resource sharing. Recently  container has become the newest virtualisation technology which solves the application isolation problem. A container allows the applications to share a host machine's operating system, making the container technology lighter and booting faster. Container technology supports two levels of sharing (software, and hardware) in contrast to the virtual machine technology that just supports hardware sharing\cite{wan2018application}. 

Container technology benefits the cloud and  virtualization with flexibility and stability\cite{wan2018application}. Containers are built on top of a host physical machine's hardware and software. Container technology allows each physical machine to have more than one container. Also, containers run images that are distributed over the cloud repositories or local hosting repositories\cite{madhumathi2018relevance}. Each container runs one application at a time. The deployment of the applications on containers is done using the container's tools. 

The deployment  on a container system is done by an algorithm called a scheduler. Also, most of the algorithms have been designed to balance the load. Container load balance enhances the container's performance. However, many other factors influence performance, such as network traffic. The dependencies between applications cause network traffic when applications are deployed in different container zones. For example, if there is more than one zone of containers in one system, and there are some applications which depend on each other, and they have been deployed in different zones, this causes network traffic. The network traffic costs the container time, and it may reduce performance. 

Our research aims to reduce and manage the traffic cost by grouping the applications with highest  dependencies in a small number of zones. Our research will investigate the background of recent technology in this field. The second part will include the most recent researches that have been done in container technology regarding load balancing and network traffic costs.    
\section{Background}
The most known container platform in the world is Docker. Docker supports cloud with Infrastructure as a Service , and Software as a Service . Moreover, many companies that invest in cloud services have been using container technology such as Amazon, Google Kubernetes Engine (GKE) and IBM cloud Kubernetes Services. 

Each container is called a cell. A zone consists of more than one cell.  Containers need a deployment system to distribute applications to  cells of zones. The deployment of applications is based upon the tasks of an application. For example, a container platform could deploy applications to cells based on load balancing to achieve the highest performance.
 For example, the round-robin algorithm has been used in the Docker platform to balance the workload, which increases the performance of the Docker platform  \cite{kaewkasi2017improvement}.

However, some algorithms have been proposed to solve other scheduling problems. For instance, the Salp Swarm algorithm (SSA) has solved the redundancy of containers at deployment time \cite{ma2019comprehensive}. Nevertheless, SSA has a problem of ``mediocre convergence rate”, which has encouraged others to propose solutions, e.g., in \cite{ma2019comprehensive}. In conclusion, a scheduling algorithm is a crucial part of a container system.

\section{Docker Architecture}
Docker was built on top of the host operation system, in contrast to  virtual machine VM that is built on the hypervisor. Hypervisor is the tier that manages the virtual machine  resources and operating systems. However, each VM should install its operating system (Figure 2.1)\cite{preeth2015evaluation}. Docker architecture has two parts that are connected. These parts are  the daemon and  the container. These two parts reside in the same host machine. Docker daemon is responsible for building and running a Docker container. Docker is based on a client-server architecture, and users use Docker interfaces (Docker client) to connect to the Docker containers\cite{preeth2015evaluation}. In addition, Docker has three main components: images, registry, and container \cite{preeth2015evaluation}.

\subsection{Docker image}
Docker image is the service that has the read-only file system that is used to build the container system. Docker uses Docker images to store and build the container system files on top of each other. For example, suppose a GO application is to be run in a Docker container. In that case firstly, the Docker image will install Ubuntu library image. After that the Docker will install the GO programming language image and the components and features that are needed to run the application. Lastly, Docker would then install the application on top of all these layers.

\subsection{Docker registry:}
It is also called Docker Hub, and it is the place where the images are stored, and from Docker registries,  Docker can upload and download the images.\cite{preeth2015evaluation}

\subsection{Docker container} 
Docker container is the place where each image and the directory needed to be run are stored. The container is the place holder for applications. The container is responsible for starting, stopping and deleting the application's images \cite{preeth2015evaluation}.  
\begin{figure}[h!]
    \begin{center}
    \centering
    \scalebox{0.5}{
    \includegraphics{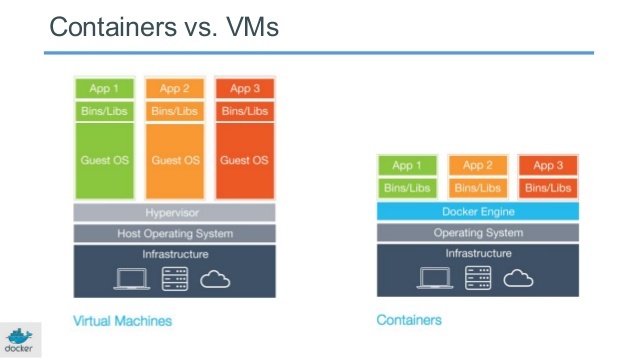}}
    \caption{Docker Container\label{fig:docker}\cite{image1}}
    \end{center}
\end{figure}

\section{Advantages of Docker}
Abdelbaky et al, 2015 explain that Docker allows users to customise and choose their applications easily \cite{abdelbaky2015docker}. Moreover,  users do not need to install either the operating system or the applications that they want to run. Docker has a significant repository of application images, and users are allowed just to clone them in their repository and use them.  Users can easily run the selected images from any distributed node  \cite{kangjin2017fid}. 

Furthermore, pulling images from Docker and running them in Docker Registry reduces bandwidth and  diminishes the time to download the images from a host node.  According to Kangjin et al., \cite{kangjin2017fid}, there are more than 150 million images hosted and run by Docker every day.  Docker allows users to deploy their applications  across different platforms and different data centres \cite{abdelbaky2015docker}. 

Although  Docker users do not have a limit on using the host machine's resources, Docker has a file system that controls resources such as CPU, shared space, and RAM \cite{preeth2015evaluation}. Even though Docker sets a limit for each resource to containers, Docker makes this limit dynamic to be increased at any time \cite{preeth2015evaluation}.

\section{Limitations of Docker}
  If the container needs to download images, and it does not have enough space on its hard drive,  it needs to download the images each time when it is required. However, after running the images, the container needs to delete them to have space in its hard disk. This process causes overhead network traffic because of  the  frequent downloading of images.\cite{ 8968919}. The redundancy of the images in the Docker cloud system is massive due to the images needed by the applications such as Apache, MySQL, and Ubuntu. 

 The security in a container system is not as strong as the virtual machine system. It is true that container technology isolates selected files of the system, but sharing CPU, RAM, and hard drive of the host machine is more likely to be compromised compared to a virtual machine\cite{8416432}. 

On the other hand, one of the essential features of cloud services is multi-infrastructure. The diverse infrastructure in the cloud has contributed to developing the dependencies between the different types of cloud infrastructures\cite{ 7850194}. Docker has developed the Swarm tool that deploys a distributed system in a multi-infrastructure network. Moreover, Swarm can develop many cloud infrastructure system dependencies\cite{ 7850194}. However, it requires three managers to ensure the availability of a Swarm cluster. If two of these managers fail, it causes the failure of the routine operation between those different cloud infrastructures\cite{ 7850194}.

\section{Container Grouping}
In this section we will introduce the mechanism of grouping resources. Docker shares the host resources with the containers that run applications. Thus, Docker has developed a tool called Cgroup to manage Docker's resources. 

Cgroup is the subsystem that runs  to manage Docker containers\cite{al2017autonomic}. Furthermore, it also produces a matrix of used CPU, RAM and I/O. Moreover, besides allocating the resources to the applications, Cgroup manages to swap resources between applications in the same type of containers\cite{7275379}. Therefore, Cgroup reduces the contention between the applications on resources. 

In addition, Cgroup minimises the traffic cost because all applications can share the same resources even though they have dependencies. However, If the applications are in the same container cluster (zone), the cost of the traffic would be meagre. The only circumstance that could make the cost of traffic high would be when the applications that have dependencies are deployed in different zones.  

\section{Container Local Contention}
Though containers allow resource sharing,  if the containers run their applications at the same time, it could cause resource contentions\cite{8758162}. For example, if there are  100 containers in the same machine that run 100 applications at the same time, all applications would share the same CPU, RAM, and I/O at the same time, hence  this will cause resources contention.  When the number of deployed containers is significant, this increases the probability of causing contention\cite{8758162}.

\subsection{Memory Contention}
Cai et al.  simulated the memory contention by deploying ten containers that ran the MySQL application. In the beginning, the applications utilised the memory gradually, but after 80 seconds from the start-up of all applications, the amount of memory utilised increased sharply. In addition, after 50 seconds from the start-up of MySQL, ten more Memcached applications started running. Therefore all the host machine memory was used up\cite{8758162}.  It has been proposed regarding this kind of situation, that moving the Memcached applications to another container would avoid the contentions \cite{8758162}. 

However, in order to limit the contention in such a situation, the resource parameters are essential, and that could be done by estimating the application's required resources before deploying them. Setting parameters for resources is good for the overall resource utilization\cite{8758162}. The  Docker parameters can be set manually but would not be precise, and that would affect the applications and system performance. Therefore, conTuner has been proposed as a framework for setting Docker parameters automatically and more accurately. conTuner ensures that all applications that share the same resources have appropriate comprehensive resources usage \cite{8758162}.
 
\subsection{I/O Contention}
McDaniel et al. \cite{mcdaniel2015two} hypothesise that to avoid contention and reach the required performance using containers can be achieved through quality of service. McDaniel et al. proposed an approach that addresses I/O contention at cluster level and node level\cite{mcdaniel2015two}. The approach is based on two layers that improve Docker Swarm for monitoring the host machine's I/O and reducing container contention on I/O. One of the layers is at node level, and the researchers assign the resource allocation to the cluster layer to ensure that each node will be given a chance to use I/O. Also, the cluster layer manages nodes while utilising a host machine's I/O. 

Moreover, McDaniel et al.  built an API allowing the client to control a container's I/O using priority. The most important feature of the API is to allow users to set the proportion of all containers for using I/O. In the cluster layer, they want to be sure that the containers will reach a high level of performance by applying Quality of Service and by monitoring all containers. Therefore, they make some changes in the Docker daemon to allow it to assign the maximum usage of the machine I/O bandwidth for each container\cite{mcdaniel2015two}. After some observation of I/O behaviour,  McDaniel et al.  modified Docker Swarm to determine the node, CPU, and RAM load, and this improved Swarm's decision of when the node can start using I/O, and when it should stop using I/O\cite{mcdaniel2015two}. The node load calculation enhanced  I/O usage by both node and cluster layers. McDaniel et al. claimed that their approach has decreased the I/O contention between the Docker's containers. 

\subsubsection{I/O Contention Based on CPU, RAM, Hard Space Usage}
Zhao et al.  proposed an algorithm that reduces the I/O contention by assigning an application to the cells where the number of applications is less or equal to the number of cells\cite{ 8259462}. Firstly, their algorithm assumes that no application will deploy unless the already-deployed applications have finished or there are more machines free of applications. Moreover, they assume that all applications have the same priority, and each application has its requirements of CPU, RAM and hard disk\cite{ 8259462}. 

The algorithm initially sorts the applications  based on I/O requirements and then assigns the application  to an available cell that has I/O available for use. The assignment algorithm considers the capability of a cell to host the application based on the application’s requirements of CPU, RAM, and hard disk as well. It then updates the status of the cell. This approach aims to reduce the local contention of resources.

\section{Container Network Contention}
Managing network traffic is an essential factor for ensuring quality of service (QoS). Docker, as a container platform, implements load balancing to satisfy the network QoS. Also, load balancing improves sharing of resources and provides an environment to deploy, implement and test applications. However, in terms of the Docker network meeting QoS standards, Docker-based on the Linux operating system uses the Round Robin algorithm. Moreover, the Docker Round Robin algorithm is used to manage network packets and to assign them to the right interfaces\cite{chuchuen2019quality}.  For example, if four packets need assigning to just two network interfaces, the first packet will be assigned to the first network interface, the second packet will be assigned to the second network interface, the third packet will be assigned to the first network interface and the last one to the second network interface\cite{chuchuen2019quality}. This mechanism ensures there is no contention between the resources in terms of using the network. 

However, in the container platform, because all applications share the same host resources and run on the same host operating system kernel, this leads to three types of network contention: cell to cell, machine to machine and cluster to cluster(zone to zone)\cite{8259462}.  Due to the increase in the availability of the network for all containers, cell to cell contention plays a small part because all the containers (cells) are in the same host machine so that the network contention will be almost zero. However, between zone to zone, the network contention is significant due to the communication between the  applications in different cells. The last contention is the machine to machine contention; this could be when applications are distributed over many machines.

\subsection{Network Contention Based on Application Dependencies}
Dependencies play a significant part in the zone to zone network contentions. The traffic cost between zones in terms of application dependencies should be minimised to enhance the network performance and reduce network contentions\cite{8259462}. Zhao et al proposed a solution to minimise the network traffic cost by calculating all the costs between one application A and all zones and finding the minimum cost and then reassigning the application based on the minimum application dependency cost regardless of the load balancing\cite{8259462}. They consider the coefficient of variance (CV) for calculating load balance, and it should be less than the user-defined threshold (UDT). Based on that, they calculate the cost of network traffic between zones. 

In addition, they attempt to minimise the cost of both network traffic and load-balance by calculating network traffic as that equals to the actual network traffic divided by the bandwidth aggregations. They do not consider the CPU, RAM, and other constraints. They define a function that maps an application to the appropriate zone. Finally, they use the arithmetic mean as the objective function which considers both network traffic and load-balance, and that makes the problem more challenging and more complicated. 

On the other hand, Zhao et al\cite{8259462} used the Traveling Salesman Problem (TSP) to represent the network traffic between zones and finding the sum of the shortest paths. They proposed that each application should only depend on the nearest two applications. However, their research considers optimising the problem that has only one application by proposing an algorithm to analyse the problem to minimise the contention at the time of deploying the application\cite{8259462}. 

\section{Container Technologies and Algorithms}

In this section, we will address the most important algorithms that contribute to container technology in load balancing and network traffic minimisation. However, since Docker is the most popular cloud container technology, we will investigate most of the algorithms that are devised for Docker or proposed to work with it. In addition, it may also include other platforms such as Kubernetes. 

\subsection{Docker Swarm}
Swarm is used to cluster and schedule containers in the Docker platform\cite{8315364}. Swarm is also a Docker management tool that was launched in 2014\cite{8446688}.  It was built to work as a FIFO algorithm meaning that Swarm selects the first job (request) and executes it. For example, if there are  new applications needing to be deployed to containers then Swarm will deploy the first application by finding the appropriate container that fits  the application in terms of container capacity of CPU, RAM and hard disk space\cite{8315364}. 

\subsubsection{Docker Swarm Strategies}
Swarm has three strategies that are used for managing Docker containers: (1) Spread (2) Binpack (3) Random\cite{8446688}. Spread allows the Docker manager to deploy the applications on the containers that have fewer applications running and this is the default unless it has been specified to the manager to deploy in a certain way\cite{8315364}. Binpack behaves in contrast to  Spread. It selects the containers (nodes) that have the most applications running. The last strategy is Random, which deploys applications on nodes randomly\cite{8446688}. Scheduling applications in Docker was considered as an NP-hard problem; however, Swarm performs well in solving the problem.  Swarm is a simple, appropriate strategy for building cloud containers and is  flexible. Swarm is one of the most popular load balancing algorithms. 

\subsubsection{Multi-Algorithm Collaboration}
Li and Fang (2017)\cite{8446688}claim that Swarm strategies would fail for scheduling jobs at specific points when: (1) none of the services of the containers are running; (2) the services require a particular type of resources; (3)  there is a significant number of requests on containers. Therefore, they developed an algorithm to solve these issues called the Multi-Algorithm Collaboration Schedule. 

Li and Fang \cite{8446688} improved the way Swarm receives requests and responds to users' requests using a different strategy. First, all client requests that ask for the creation of a container will be held when Swarm is busy performing other tasks. Second, the data of the node that is currently running will be collected by DATA Gather. Third, schedule requests the DATA Gather for cluster load status. They called their strategy ``Dispatch Module", and the algorithm can add the application in the queue based on the its requirements of a machine's resources. Li \& Fang \cite{8446688} observe that their strategy has a better outcome in terms of load balancing compared to the Swarm strategy.

\subsubsection{Docker Swarm Strategy Based on Service Level Agreement (SLA)}
C'erin et al. \cite{8315364} proposed a strategy that adapts with Swarm in three stages in terms of creating a container to deploy an application (1) Container Scheduling, that is selecting request from a "user requests pool"; (2) aggregates the information of the containers that could adapt the applications(the containers satisfies applications with enough resources) this stage is called Resources Allocation; (3) the last stage selects the container that runs the application\cite{8315364}. However, C'erin et al. \cite{8315364} claim that the significant problem with Swarm is that Swarm cannot accept  more than one request at the same time\cite{8315364}. Therefore, they implement their strategy by considering the priority of the requests. The strategy aims to allocate applications to containers as soon as the calculation of the resources is completed. 

However, they expect that the algorithm will stop all processing when an individual request that has high priority does not fit in any container. Even though there are containers available that could run low-priority requests in the queue, therefore, the waiting time for the low-priority requests will increase. Their proposed strategy aims to recalculate the resources and allocate the appropriate request that fits within a container so that Swarm will continue serving the queue with requests that can be run on an available machine. The priority in this model is based on three types of classes which are the Premium, Advanced, and Best effort, and all of them are based on the client's budget\cite{8315364}. Premium is the highest priority, then Advanced, and the lowest priority is Best effort. The strategy selects the container based on these priorities; then, with dynamic resources, it allocates the container to the machine\cite{8315364}.  

\subsection{Kubernetes}
One of the most widely known of container orchestration is Kubernetes that runs the services inside containers automatically and with flexibility. Kubernetes performs load balance, fault tolerance, and scales the job horizontally\cite{beltre2019enabling}. In addition, it supports Docker containers, and it has a set of container management tools called "Pods" which is also used in the Docker platform. However, Kubernetes has requirements to support MPI that run HPC applications which is responsible for  inter-container communication, and has access to the hardware\cite{beltre2019enabling}. 

\subsubsection {Kubernetes Infrastructure an Components}
Kubernetes has two infrastructure components which are Master and Node\cite{beltre2019enabling}. Kubernetes Master is responsible for making decisions that will be made at the cluster level. According to Belter et al \cite{beltre2019enabling} Kubernetes Master has four components which are Kube-controller-manager, Etcd,  Kube-scheduler, and Kube-apiserver. Furthermore, each of these components has specific functions. For example, Etcd is responsible for saving all cluster information and the backup. Also, Kube-scheduler is responsible for assigning Pods to  resources based on the  requirements of the Pods. 

Moreover, admin of Kubernetes gains access to the Kubernetes Master through API server.  The API server is a middleware layer between the Kubernetes master and the nodes. Figure 2.2 Shows the Kubernetes components and communications. In addition, Kubernetes needs three components to run, that are Kublt daemon, a proxy for communication, and container run-time.
\begin{figure}
    \begin{center}
    \scalebox{0.5}{
    \includegraphics{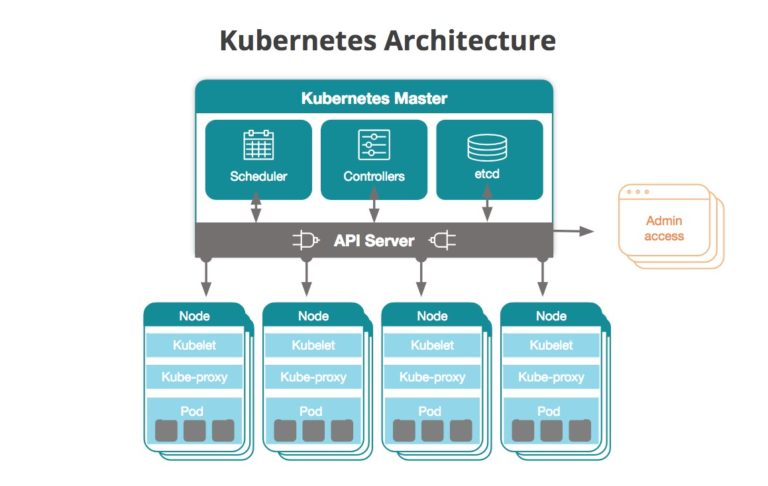}
    }
    \caption{kubernetes Architecture\cite{image2}\label{fig:docker}}
    \end{center}
\end{figure}

\subsubsection{Kubernetes Load-Balancing and Traffic Latency}
Load balancing in Kubernetes is crucial as well as communication between users, and service through the network. Fahs and Pierre\cite{8752909} described that Kubernetes' communication between the end-user and the services is done as a request through an IP address. The request will route to Kubernetes nodes, and the probability of each node getting the request in Kubernetes is equal. Therefore, the load balancing of jobs in the node will be high.  Moreover, Kubernetes communication (routing) between a user request and Pods has two steps (1) from an external user as requests to the Kubernetes's node, and (2) from Kubernetes manager to internal applications (Pods) \cite{8752909}.

The first step of network traffic, could be from any request to any Kubernetes node as the destination or getaway node. The second traffic step is implemented inside the Kubernetes nodes, and it is the daemon's responsibility to call Kube-proxy to route the request to the node. Kube-proxy has the route table, and its responsibility is to update it and manage the load balance for all nodes. Kube-proxy loads the balance by create the rules for all nodes\cite{8752909}. The internal routing has two main priorities. First routing is between the user and Kubernetes master through a user request and the second is routing user's request from Kubernetes master to the Pod. In order to reduce the traffic latency in the internal routing, Fahs and Pierre\cite{8752909} redesigned the internal routing and they called their module "proxy-mity". Proxy-mity allows the Kubernetes administrator to manage the routing between a user's requests and the pods while balancing the load on the pods, and that is by increasing the proportion of routing. Proxy-mity could change the load balance to reduce traffic latency and keeps it low. 

\section{Container Load Balancing and Network Traffic Algorithms}
\subsection{Container Load Balancing Scaling Algorithms }
The container Media Service uses load-balancing to reduce the waiting time, but the load-balance is independent of network traffic cost \cite{10.1145/3022227.3022243}. Kim et al\cite{10.1145/3022227.3022243} suggested an analyser that analyses the container traffic called (COTA). The analyser aims to ensure that the container will be available. Moreover, they developed an algorithm which is Least Traffic Load Balancing (LTLB) for balancing the load among containers. The LTLB algorithm monitors the traffic between the containers\cite{10.1145/3022227.3022243}. They also suggested an auto-scaling container algorithm to enhance the container service performance in terms of rapid access by users. 

 They\cite{10.1145/3022227.3022243} also considered a dynamic load balancing algorithms such as Least Connection algorithm (LC), and a WLC algorithm instead of s static load balancing algorithm such as the Round-Robin (RR) algorithm. Round-Robin (RR) algorithm does not consider the traffic scale. To reduce and manage the traffic problem, they let all servers report the information of network traffic. However, they divided their proposed auto-scaling algorithm into two types: Reactive and Proactive auto-scaling. 

The Reactive scale is a provider's policy that ensures that all resources are used evenly to manage the load balancing. Moreover, Reactive scale monitors CPU, RAM, and I/O utilisation. Reactive scale is a service-side responsibility. The Proactive auto-scaling predicts the next step by monitoring resources such as CPU, RAM, and I/O. Both Reactive and Proactive are used for auto-scaling in VM. 

\subsection{Container Network Traffic Algorithms }
An auto-scaling method for cloud container  has been proposed by Kim et al. \cite{10.1145/3022227.3022243}. 
They proposed this method of scaling because, in the container environment, the users connect to the container service for a long time. The long period of connecting requires high performance while considering the network traffic.

 Kim et al. \cite{10.1145/3022227.3022243} proposed an analyser called (COTA) on the server-side with some roles which are: (1) monitoring the traffic collected rate by Monitoring Aggregator. Monitoring Aggregator collects the data from the server and based on the traffic information that sent to the container; (2)a violent report will be reported by Reporter to the manager; (3)  Load Balancer will use data supplied by Container "Scaler". The COTA algorithm  enhances virtual machine auto-scale and assists the load balancing algorithm. They conclude that the dynamic load balancing algorithm is like Least Connection in being more effective in terms of balancing the network utilization\cite{10.1145/3022227.3022243}. 
 
 \subsection{ Algorithm for Optimizing Container in Multi-Objects}
Liu et al\cite{liu2018new} developed an algorithm for multi-object to improve the container performance. Their algorithm aims to assign the new container to the appropriate node. They consider all containers that have the same service to be  at the same node. Each node needs sufficient resources to run the service such as CPU, RAM and I/O. They assume that cantered container schedule reduces the consumption of network traffic and increase container performance. Therefore, they propose an algorithm with multi-objective such as improving CPU, and RAM for nodes' usage as well as  the traffic time to download the container’s image\cite{liu2018new}.

Their algorithm considers the node as a set of containers that are grouped. When a new container is available to be allocated to the node, some factors must be calculated to ensure the service quality.  The service quality is determined based on the business service perspective.  However, Liu et al.  \cite{liu2018new} add more factors into the Swarm strategy in terms of enhancing the server quality by reducing the network traffic  used when images are downloaded from Docker registry, the communication between the nodes and their containers, and improving the clustering of the containers. They had used tools to measure the CPU, and RAM usage which calculate the ratio of the unused CPU and the RAM. Also, Liu et al \cite{liu2018new} had to design a tool to calculate the traffic performance and timing when the user wants to download an image from a Docker container to the local registry.  On the other hand, Docker Swarm helped them to cluster the containers in nodes.  Finally, they store these factors and information in a data structure, which also has the scores of nodes, and they are sorted accordingly. 

The algorithm by  Liu et al\cite{liu2018new}  examines the usage of CPU, RAM, communication between nodes, clustering of containers, and the network traffic time that may be needed to pull an image when the new container is allocated. Indeed, the main idea of their algorithm is to find the best place to allocate the new container among the nodes.  However, they aim to cluster the nodes based on the five factors that have been proposed. Therefore, when they compared their algorithm with Swarm strategies, they found that their algorithm performs similar to the  Swarm strategies. Finally, Liu et al \cite{liu2018new} concluded that the connection that could be built between nodes and their containers and the way of clustering the containers together increase the efficiency of the scheduling process in terms of using the resources such as CPU and  reduced the network traffic.  

\section{Conclusion}

The cloud container technology is the most lightweight virtualization technology for a host and runs the applications on distributed systems. Cloud container is the environment that deploys applications and isolates them in a small part of the host machine resource. Also, container technology enhances the sharing of the host resources. Although it is the most recent technology, it has been an active field of study and research in recent years. The most important aspect of container technology that the researchers have addressed is container performance.  

The gaps in this field that most of the research does not consider is the dependencies between the applications. The dependencies between applications that may be deployed in  different containers cause network traffic. Network traffic reduces the performance of the containers.  While some research does consider the contention of resources such as  contention for local and network resources,  the proposed algorithms either solve this problem by putting all the applications into the same zone or controlling the host resources.

 Therefore, our research is concentrated in developing an algorithm that deploys applications on containers based on the application dependencies and load balancing. Our research also considers the cost of network traffic  that is caused by the  dependencies before and after applying the proposed algorithm to illustrate the reduction of the network traffic costs.

\chapter{Definitions and Research Objectives}
This research aims to improve  container performance by designing an algorithm that reduces the network traffic and balances the loads  of containers. To achieve this goal our algorithm considers four aspects which are: 
\begin{enumerate}
\setcounter{enumi}{0}
\item The application resource requirements from the host  machine. Also, the host's capabilities during the deployment and how it may influence the load balancing.
\item The application dependencies and network traffic overhead.
\item The overhead of moving applications between containers. 
\item The container performance affected by load balancing and network traffic overhead.
\end{enumerate}
However, to disclose the objectives of this research, the aim is to understand the problems first and at the end of this section, the research objectives will be discussed. 

\section{Application Requirements and Host Resource Capabilities}
In order to run any application, the application needs to meet at least the minimum of its resource requirements such as CPU cores, RAM size, and the minimum amount of hard disk. On the other hand, each host machine offers a specific amount of resources that has the potential of satisfying the application requirements. However, challenges emerge when the applications are starting to be deployed; the schedule should deploy the applications to an appropriate host machine. This step ensures that the resources will be utilised efficiently. 

\section{ Network Traffic Overhead}
Network traffic  overhead influences the performance of  cloud containers, due to the dependencies between the applications. Furthermore, extreme use of network traffic may cause a delay of the  execution times of applications, and it may cause increase in network latency in contrast to independent applications that do not need to use network traffic. Moreover, dependent applications that are in the same zone cost almost no  network traffic. 

\section{The Overhead of Moving Applications Between Containers}
Moving applications from one container to another has an additional cost of network traffic, but it is a one-time cost. The cost of moving applications between zones could cost ten times more than each application's request or response. Despite the cost of moving applications, the accumulative cost of application requests could be higher than moving applications, mostly when the applications use network traffic for a long time. 

\section{Container Performance and Network Traffic Based on Applications Dependencies} 
Lower network traffic is responsible for reducing the latency, and it increases application deployment into the containers. Essentially the container systems are responsible for running applications using their host's resources, and container system should deploy the applications equally on containers to increase the performance. Also, reducing network traffic works in addition to load balancing to boost  performance. Although the dependencies between application do not affect the load balancing, a high volume of network traffic delays the distribution of the applications.

\section{Research Objectives} 
This research aims to demonstrate that  load balancing and network traffic influence  container performance, especially when the dependent applications are deployed into different zones. Moreover, the research aims to develop and examine an algorithm to solve this problem. However, the research objectives will be broken down into three main objectives. 

The first objective is to examine containers by deploying dependent applications on different container zones using a modified Round-Robin algorithm. At this stage, the research will not consider the dependencies between the applications. The goal is to measure the network traffic cost and   the load balancing of  containers in different zones. Furthermore, this will illustrate the container performance based on load balancing and the overhead of using network traffic. 

The second objective is to develop an algorithm that reduces the network traffic by reducing the  dependencies between the applications. This research considers this problem as the Maximum Cut problem, which means that the algorithm should group the applications based on dependencies. 

To illustrate how the algorithm behaviour can solve the maximum cut problem,  figure 3.1 shows that in cut 1 the applications are divided into two zones of containers. The applications 1 to 5 are distributed equally into the zones. Moreover, the figure shows the applications as the vertices and the dependencies as the edges between vertices in a graph. Cut (2) shows the first attempt to cut the dependencies between zones, by  moving application 1 to the left zone. Furthermore, the last step cuts dependencies by 40 per cent. In cut (3), by moving application 2 to the right zone, this step does not cut any dependencies at all. However, cut (4) reduces the dependency by 60 per cent by moving application four from the right zone to the left-hand zone, and that is the maximum cut so far. The algorithm applies a straightforward methodology to reduce dependencies. Also, our research aims to measure the reduction of network traffic cost.  The total cost of using the network traffic for the period as well as moving the applications between zones will be measured. 
\begin{figure}[ht]
    \begin{center}
    \scalebox{0.38}{
    \includegraphics{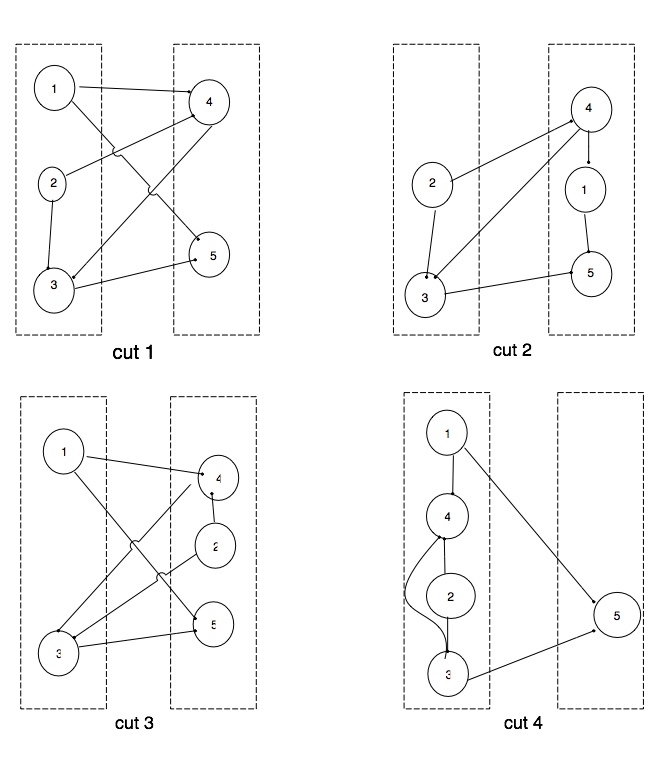}
    }
    \caption{Max Cut Illustration\label{fig:maxcut}}
    \end{center}
\end{figure}

The idea of using the maximum cut problem in this research is to find the best cut that reduces the network traffic and to group the applications in zones based on their dependencies.  

Finally, because the second objective does not consider load balancing, the last objective is to enhance  load balancing. to conclude, as most research do not consider dependencies between application during applications deployment, the objective is to develop an optimum algorithm that reduces the network traffic usage and managing load  balancing at the same time. By applying these two objectives it is hypothesized  that the container performance will be boosted.

\chapter{Methodology}
\label{appa}

The methodology is divided into two main sections.  These are the algorithm description, and algorithm design. In this chapter, we will attempt to explain the ways the algorithm was developed and the challenges associated with implementing the algorithm. 

\section{Algorithm Description and Design}

The objective of this research is to develop an algorithm that increases the performance of cloud container systems by  balancing the workload and reducing the network traffic between the dependent applications that may be deployed in different zones. Firstly, terminology and key words that are used in this section will be introduced.  Secondly, we will explain the algorithm specifications as related to the algorithm design.    
\subsection{Terminology and Key words }
In this section, an explanation of the most frequently used terminologies.
\begin{enumerate}
\setcounter{enumi}{0}
\item \textbf{Container:} is the smallest component of cloud container systems. It includes a bundle of  software code. This research refers to the container as the place that hosts the application, and all software layers that might be needed to run  applications such as operating system files and programming language compilers. 

\item \textbf{ Zone :}  also called a Pod, which represents the group of containers that share a common environment components, such as the CPU, RAM, and hard disk. This research will use the zone as the keyword that describes the containers that are grouped in the same Pod. 
\item \textbf{Node:} is the machine that runs the zones. However, the machine could be physical or virtual machine. 
\item \textbf{Cluster:} is the group of nodes regardless of the infrastructure of the nodes. 

\item \textbf{Network Traffic Cost:} refers to all the cost of using the network. The network traffic cost will be considered as the following:   
\begin{enumerate}
\item \textit{ Traffic inside the zone} is the network traffic cost, for applications using the network in the same zone and that costs  10 time units for each network communication\cite{RN13}. 
\item \textit{ Traffic between the zones} is the network traffic cost between the zones and the cost would be 10 time higher than traffic inside the zone \cite{RN13}. 

 \item \textit{Moving application between the zones} is the network traffic cost of moving one application between the zones and the cost would be 10 time higher than traffic between the zones.
\end{enumerate}
\item \textbf{Graph} represents the connection of dependent applications that are deployed on zones. The vertices represent the applications while the edges are the dependencies.

\end{enumerate}

This research will consider\textit{C} to  denote a container, \textit{Z} to denote a zone, \textit{A} to denote an  application and \textit{T} to denote the network traffic cost. In this research, each zone is assumed to have its own resources.  For example, `zone0 3 CPU' means that zone0 is equipped with a CPU that has three cores. Also, `zone1 8 RAM' means that zone1 has 8 GB of RAM. \textit{$A_{cpu}$} denotes how much the application requires from CPU. Lastly,\textit{$D_{i,j}$} refers to the dependencies between  applications $i$ and $j$.

\section{Algorithm Specifications}
The algorithm has been divided into three approaches to meet the objectives of this research.
The first approach is to evaluate container performance by only considering the  balancing of the workloads. The second approach is to evaluate container performance by only considering network traffic. The last approach considers both load balancing and traffic, and the presented algorithm will compare the  results of the three approaches to illustrate the differences. We will also evaluate the extent to which  these factors affect the performance of the cloud container. 
\begin{figure}[h]
    \centering
    \scalebox{0.40}{
    \includegraphics{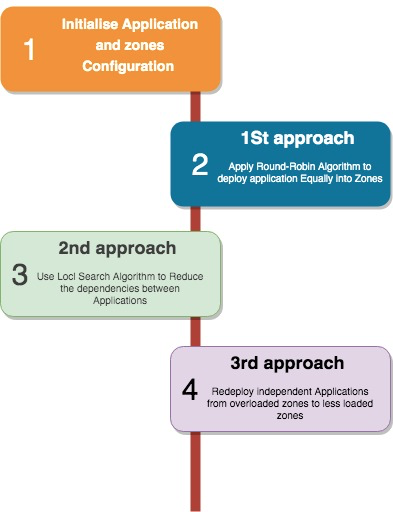}
    }
    \caption{System design}
    \label{fig:my_label}
\end{figure}

The algorithms have been simulated  and runs the three-approaches respectively as shown on Figure 4.1,  ``System design”. The main steps of the simulation are: 
\begin{enumerate}
\setcounter{enumi}{0}
\item Aggregating the application specifications from the user and getting zone specifications from the cloud's configuration files.
\item Sending the information to the schedulers. The scheduler deploys the applications into zones. 
\item Showing the result and statistics. 
\end{enumerate}

The rest of this chapter will explain the three approaches. 
\subsection{Deploying Applications into Containers Based on Load Balancing}
In this approach, the aim is to enhance the container performance by balancing the workload of the cloud container to ensure that the experiment is realistic. Therefore, it is assumed that each zone has almost the same number of containers. Moreover, each zone has specific resource capacity. Also, the applications must have minimum requirements of resources.

A Round-Robin algorithm for managing and deploying the applications into zones will be implemented. The algorithm was developed to satisfy two parameters which are the load balancing and the resource requirements of applications in zones such as, CPU, RAM, and hard disk. 
This algorithm assumes that all applications will arrive at the same time. Firstly the algorithm sorts  zones based on capacity. For example, the lowest capacity of resources of the zones would be zone 0. This step reduces the iteration of the algorithm and finds the appropriate zone for an application more efficiently.

\begin{figure}[h]
    \begin{center}
    \scalebox{0.39}{
    \includegraphics{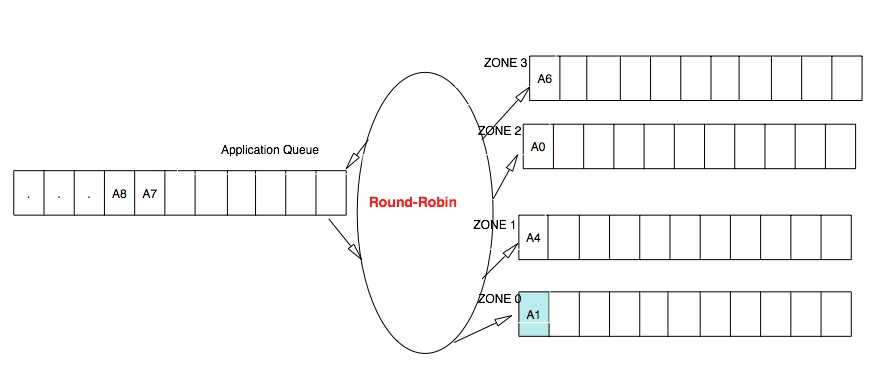}
    }
    \caption{Round Robin Algorithm\label{fig:RoundRobin}}
    \end{center}
\end{figure}  

Secondly, after sorting the zones, the developed Round Robin algorithm puts all applications in a queue. The first round of the Round Robin algorithm selects the first application, and it will assign the application into the zones that have equal or have more resources than the application requirements of resources (Figure 4.2). 
We believe that The Round-Robin algorithm deploys the applications equally into the zones. 
This approach will boost the container's performance in terms of response time, application availability and flexibility. Nevertheless, the network latency would be high due to the dependencies between the applications.

On the other hand, to measure the load balance a specific equation that calculates the load balancing and gives a clear indication of the load balancing situation is needed. Zhao et.al \cite{8259462} used adjusted coefficient variation (CV) to measure the load balancing and  coefficient variation returns  almost the same number for  different load balancing situations. The best practice of load balance is when each zone has the same number of deployed applications. In this situation the CV will be zero.  The following equation calculates the CV for measuring load balancing.
\begin{equation}
    \begin{aligned}
         CV =\frac{\sigma }{\mu }\\
         where 
    \end{aligned}
\end{equation}
\begin{equation}
    \begin{aligned}
\sigma =\sqrt{\frac{1}{\left | z \right |}\sum_{i=0}^{n}(A_{z,n}-\mu )^2}\\ where \\ A_{z,n}=Application\ number\ in \ zone \\ and
\end{aligned}
\end{equation}
\begin{equation}
    \begin{aligned}
\mu  =\frac{\sum_{i=0}^{n}A_{i}}{\left | Z \right |}
\end{aligned}
\end{equation}

Due to the difficulties in determining the worst indicator number for the worst load balancing and comparing  with the  previous equation, the next empirical equation considers the worst case of load balancing when all applications are deployed in a single zone. In this situation, CV would be equal to the square root of the number of zones. This equation will measure the percentage of the load balancing more accurately, and it will provide a clearer indication of  load balancing.  
\begin{equation}
    \begin{aligned}
    Load Balance=\frac{100-(100*CV)}{\sqrt{\left | Z \right| }}
\end{aligned}
\end{equation}

\subsection{Deploying Applications into Containers Based on Network Traffic Cost}
Due to the network traffic cost as a cost of network latency, lost packet, and queue delay,  
the network traffic plays a significant part in container technology.  Amazon states that network performance has a significant impact on the services of the cloud \cite{RN6}. Therefore, in this approach, we aim to reduce the dependencies between applications, analyse the output, and compare them with the output of the first approach to develop the last approach.  

This approach continues from the previous approach. It redeploys the applications into the zones, based on the application dependencies. The container architecture is considered as a graph in which the applications are the vertices, and the dependencies are edges. The dependencies between the applications are thus divided into two types which are:
\begin{enumerate}
    \item \textit{ Internal Dependencies,} which are the dependencies between applications that are in the same zone. 
    \item \textit{External  dependencies,} which are the dependencies between applications that are in different zones. 
\end{enumerate}
This approach aims to reduce the external dependencies, which are more expensive than the internal dependencies. Figure 3.1 shows these two types of dependencies. Through aiming to redeploy the application by considering the Maximum Cut problem, a modified Local Search Algorithm will be applied to reduce the network traffic between zones as it groups dependent applications in the least number of groups. However, the applications that have both external and internal dependencies, it would be no benefit to move them from their zones. The Local Search algorithm, however, solves this problem as it considers the following steps: 
\begin{enumerate}
    \item Calculating the number of external dependencies.
    \item Selecting the application that depends on  applications in other zones or services.
    \item Finding the neighbours of this application. Finding the application's neighbour is based on the highest network traffic cost.
    \item Moving the application to the application's neighbour's zone. Moving an application between zones cost 10X of the network traffic cost, thus moving applications between zones cost (1000 units).
    \item Recalculating the number of external dependencies. If the number of dependencies is reduced, the algorithm will commit the step and select the next application, but if the number of dependencies has not decreased, the algorithm must return the application to its original zone and then it would select the next application.  
\end{enumerate}

In this stage, it is assumed that this algorithm will reduce the dependencies between applications. Thus, it reduces the network traffic cost. The algorithm enhances the performance of the container by reducing the latency, reducing the executing time of the applications as well as the response time. However, because the load balancing in this approach is not considered, it is expected that the  workload of the zones will be unbalanced. 

Although the traffic between applications in the container offers availability of the services, it increases network congestion. In addition, packet loss and paralysis occur if the path of the network is crowded, and this would also cause disconnection of communication between applications and system failure \cite{8968700}. Therefore, to avoid losing packets or path failure more than one path between resource containers and destination containers or services is needed. The goal of having redundant paths in the network is to solve the traffic issues in the network, as considered earlier. The redundant paths guarantee the availability of the services and increase the performance of containers. Therefore, Fat-Tree  as a network architecture is assumed in this experimental network topology. 

The Fat-Tree is the most widely used topology for multi-paths in the network. It allows a tremendous amount of network traffic at low cost, and it utilises the paths sufficiently \cite{duan2015cost}. Cloud container applications need a significant number of packets that must be transmitted through the network\cite{duan2015cost}. Moreover, most of the packets are of the same types, including requests, queries, and service response, which may cause heavy network traffic, packet loss, and  latency\cite{duan2015cost}. Thus, the Fat-Tree topology of the network is the appropriate solution for solving this problem. Most of the systems use a Fat-Tree that has three tiers of switches. These switches route the packets through the network and provide the network with multi-paths\cite{8968700}. The three tiers are core tier, aggregation tier, and pod tier. Figure 4.3 shows the structure of the tiers. 
\begin{figure}[h!]
    \begin{center}
    \scalebox{0.20}{
    \includegraphics{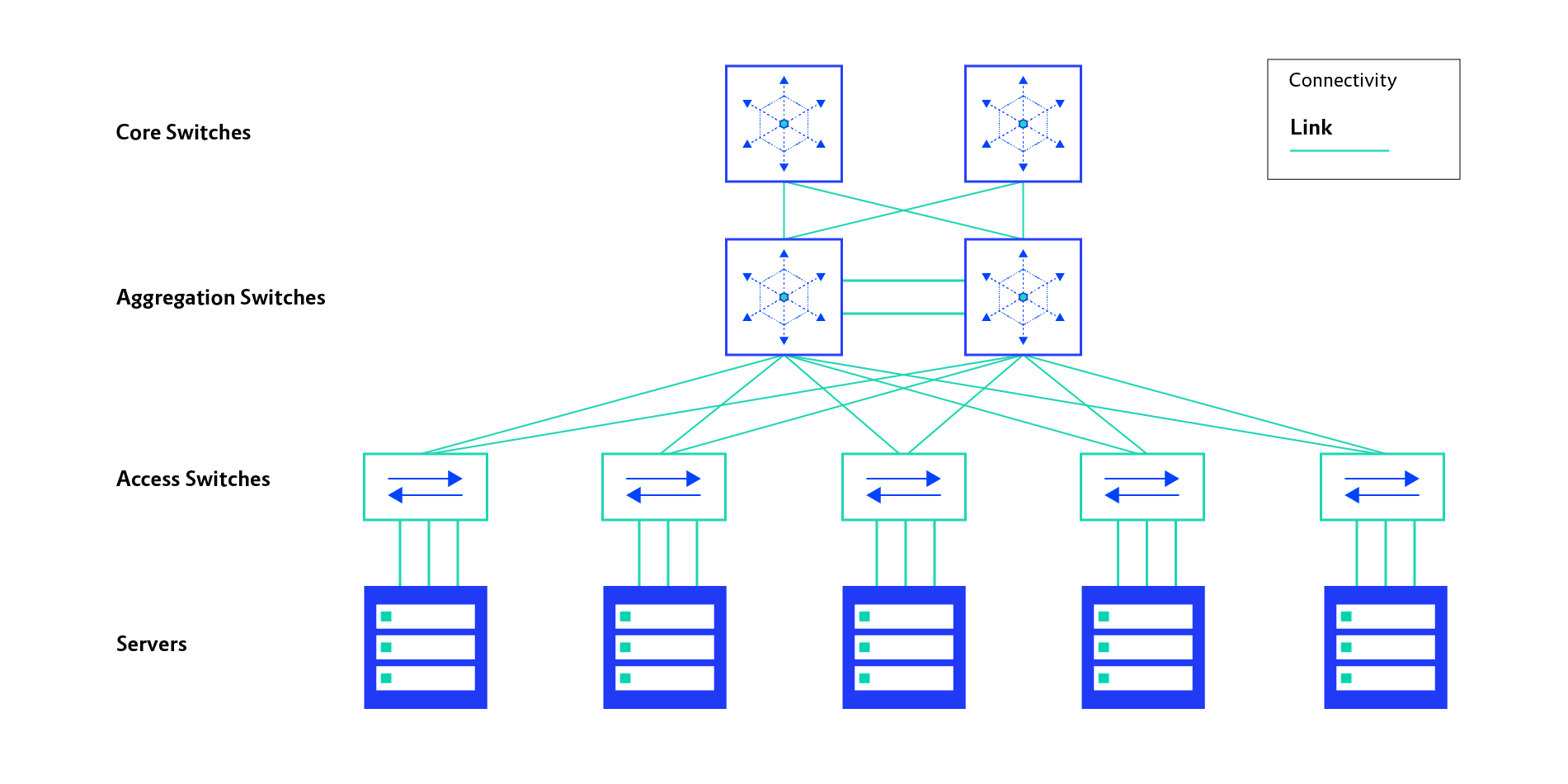}
    }
    \caption{Fat Tree Infrastructure\label{fig:FatTree}}
    \end{center}
\end{figure}  
The following explains the routing infrastructure of the tiers: 
\begin{enumerate}
    \item \textbf{Pod  Tier} is the layer of switches that routes the packet between containers inside the node(server). These switches may be called access switches. The number of switches in this layer would be half of the number of containers. Each switch connects two containers.  
    \item  \textbf{Aggregation Tier} is the layer of switches that routes the traffic inside the zone. This layer allows the packets to reach the pod tier with more than one path. The number of switches in this layer is the number of pod's switches in the same zone. 
    \item \textbf{Core Tier } is the layer of switches that is at the top which connects all the aggregation switches and routes the packets between zones. It works like a gateway, and it has the same number of switches as the number of the zones.  
\end{enumerate}

The research that measures the latency in a network \cite{guo2015pingmesh} shows that the maximum overall latency in a data centre DC for one day is 1397.63ms. It also shows that the maximum latency for one DC is 0.268ms in a zone and 1.34ms between zones. Also, Zhou et al described that the network traffic cost between zones is 10 times larger than traffic cost within a zone. They argued that the latency of Amazon EC2 within North California was 34 ms and between Singapore and North California was 385ms. Thus, they assumed that network traffic is 10 in the same zone and 100 between zones\cite{guo2015pingmesh}\cite{8259462}\cite{RN13}. Therefore  Zhou et al had built their traffic cost matrix based on their assumption\cite{8259462}. Hence we have based our costs also in a similar way. the latency (cost) between the zones is ten times larger than the latency inside the zone. Based on this, to measure traffic cost of the network, we consider there are three types of costs (1) the cost of traffic inside the pod tier, which costs 10 time units for each edge; (2) the traffic between the nodes through the aggregation tier and core switch tier costs 100 for each edge; (3) It is considered that moving an application between containers costs ten times more than  a request or response packet because of the size of the application. If, however,  the application \textit{A1} in zone 0 sends a request to a service or application \textit{A2} in zone 4 and the request will go through two aggregation switches and one core switch to zone four, the packet will go through $E$ edges and this costs the number of edges $E$ multiplied by  the cost $c$ of each edge, i.e., $Ec$. Therefore, the cost of each request is based on the routing path, and to get the optimised cost, the shortest paths must be found.

It is assumed that the topology of the network is a Fat-Tree because the Fat-Tree guarantees the availability of the service and it is fault-tolerant. As the fat-tree has three levels of the switch for transferring the data through the network, the traffic calculation is to aggregate the number of  edges between the source container and destination container. 

Thus, if there are two applications in the same zone, but they do not connect to the same switch, they need to go through the upper level of the network switch, which is the aggregation level. therefore, the cost would be higher than if they share the same switch.

The three Fat-Tree levels of switches connect with S number of switches and K number of ports. For example, if there are N number of pods, which in this case are the containers, the number of switches would be N/2 for both pod layers and aggregation layers and, the number of switches in the core layer is the number of the zones divided by two times (Z/2), and each switch has port for each zone \cite{8968700}.

As the Fat-Tree allows the multi-path between sources and designations, there is more than one path that connects containers. Each path costs a certain amount based on the number of the path edges. If the traffic either for requesting, responding, or moving application takes the shortest path between the source and destination containers, the previous formula that calculates the cost of traffic would be modified.  This modified formula is to calculate the edges $D_i,_j$ based on two parameters, which are the level of the edge and the cost of the edge. L was denoted as the level of the network and \textit{T} as the cost of the edge, while \textit{Li} as the number of edges on different tiers.  

\subsection{Combining Load Balancing and Network Traffic Algorithm}

This approach aims to enhance the  balancing of workloads that has been affected by reducing dependencies. It is an additional step to reach optimum performance by combining the two previous approaches. Thus, to continue from the second approach and redeploy the applications into the zones to improve the load balancing, the steps of this approach are the following:  
\begin{enumerate}
    \item Finding an independent applications in the overloaded zones;
    \item Moving the independent applications to an under-loaded zones that the selected application can fit into;
    \item This iteration will stop when there is no under-loaded zone that can host the selected application; 
\item Starting the step over for the next application;
\item The approach will stop moving independent applications when the load balancing reaches 100\% or when there is no more independent applications in overloaded zones.
\end{enumerate}

This step will increase the network traffic by moving applications between zones, but it enhances the load balancing. Indeed this approach is the combination of the previous approaches, and when the algorithm is divided into three approaches, it illustrates the collaboration between the approaches  to enhance  container performance. By grouping them and considering the load balancing, the performance of the container system increases.   
\section{Algorithm Design } 
When the algorithm pieces are collected, they give the reader the complete picture of the steps of the algorithm. Although the algorithm seems that it contains three approaches, in fact,  it is an incremental algorithm. The following steps are involved in the algorithm design: 
\begin{enumerate}
    \item Sorting the zones  based on each  zone's capacity. 
    \item Using CPU, RAM, hard disk as constraints for both applications and zones. 
    \item Using the Round-Robin algorithm to assign  applications to  zones based on the constraints. 
    \item Calculating and printing the cost of using the network traffic. 
    \item Calculating and printing the load balancing indicator for zones.
    \item Collecting and printing the data for container performance. 
    \item Reducing the application dependencies by considering  as a maximum cut problem, this reduces the network traffic between zones.
    \item Repeating steps 4-6. 
    \item Redeploying independent applications from overloaded zones to the appropriate under-loaded zones.
    \item Repeating steps 4-6. 
\end{enumerate} 
It is assumed that this algorithm illustrates the progress of the improved performance in addition to achieving the research goal and answering the research question.  The Algorithms 1 shows functionalities:  
\section{The Algorithm}
\begin{algorithm}
\KwData {input: dependent applications A1.....An and ascending sort: the zones based on the capacity }
\KwResult{ deploy all applications to zones' containers $C_{z1}....C_{zn}$, reduce the network traffic and enhance the load balancing}
  \While{ Round-Robin Algorithm deployed all application==false}{
  deploy all application to the zones equally }
calculate the costs()\{\\
  \While{all zones}{
      calculate mean, Standard Derivation(SD),Coefficient Variation (CV)\;
    }
  \While{all applications}{
      calculate trafiic cost (tc)\;
  }
  \}\\
  \While{all applications and zones}{
     
       $count \gets count$ (dependencies) 
         find the neighboured(A1)=Am 
         using Local-Search algorithm to remapping ( $zones\gets applications$)
      $ Newcount \gets count(dependencies)$ 
      \eIf{Newcount$<$ count \&\& $A_{Requirment}<Z_{capacity}$}{
            $C_{z,Am}\gets A1$
         }{
       $A_{i++}$
         }
    } 
 back to calculate the costs:
 \While{independent application}{
     \eIf { $A_i$ deployed in overload zone==true}{
             \eIf {$Z_i$ == empty container \&\& $A_{Requirement}<Z_{capacity}$}{
                                 $Z_i\gets A_i$
                    }
                         {
                                  $ Z_i$++
                         }
         
        }
        
    }
go back to calculate the costs:
\caption{Research Algorithm Design}
\end{algorithm}
\subsection{Container Performance}
This section demonstrates the containers' improved performance based on two factors that are load balancing and network traffic. The relationship between the load balancing and cloud containers is a linear relationship which means that as the load balance becomes optimum, the performance will become better. In this research, the load balancing has been calculated as a coefficient variation (CV).  This means the best case for load balancing is when the CV equals zero.

However, the relationship between network traffic and performance is in inverse relationship, and the most reliable performance in cloud container occurs when the network traffic is almost zero. As it cannot be guaranteed that the network traffic will be zero, the best case is when all applications are independent, or all dependent applications are in the same zone. The function traffic \textit{f(tr)} indicates the relationship between containers performance ,and the network traffic cost and the load balancing of the zones. Equation 4.5 shows the container performance.

\begin{equation}
\begin{aligned}
F(Per)=\frac{\alpha +\beta }{f(tr)+\lim_{CV=0}f(cv)}\\ \\
where \ \alpha \ is \ constant \ factor\ of \ traffic \ \&\\ \\ \beta \ is \ constant \ factor\ of \ load balancing 
\end{aligned}
\end{equation}

\chapter{The Experiment}
The algorithm was implemented by simulating the cloud container deployment system. GO programming language was used to develop the simulation. The simulation was run on the Mac operating system with hardware capacity of a CPU of 2 GHz Intel Core i7 and 4 GB of RAM. More than ten packages were developed for the simulation. The packages handled the initialising of the simulator with applications and zone data and run the algorithm for deploying applications by the three proposed approaches. Also, the simulator showed the results and statistics.

The simulator processed and implemented the proposed algorithm in three stages, and each stage had to implement several processes and tasks.  Firstly, the algorithm simulation initialised the experiment by reading the experiment's file configurations. Experiment file configurations include all information of the application requirements and zone capacities of resources. These configuration files also specify the experiment scale, traffic cost and the dependencies between applications. \textit{Appplicationdis,zone}, and \textit{interfacedis} packages were responsible for initializing the simulation and reading the configuration files. 

The simulator initialised the \textit{ApplProperty, ZoneProperity} classes that contain the applications and zones names, application requirements, and zone capacities. Moreover, in this stage, the simulator created a dependency matrix, traffic cost matrix, a queue of applications and groups of zones that will host the applications. 

The first stage after the initialization step was to pass all the information to the \textit{roundrobingroup} package. This package applies the modified Round-Robin algorithm to deploy the applications into the  containers of zones. \textit{GroupingApp()} is the function that is responsible for mapping the application to the  containers. Figure 4.1 shows how the \textit{GroupingApp()} works. The second step in this stage was passing the outputs of this step, which is the deployment of applications into the zones to \textit{calculatedepen} package that calculates the following 
\begin{enumerate}
    \item Counting of the external dependencies and internal dependency by the \textit{CalculateDepFunc()} function.
    \item  Calculating the traffic cost by the \textit{CaculateProprotionOFtraffic()} function. 
\end{enumerate}
In the third step  the simulator invokes the \textit{stitic} package to calculate the load balancing and cloud performance by   \textit{CountAppInEachZone(), CalculateMean(), CalculateSD(), and ColudPerformance()} functions. However, the last two steps were repeated in all stages.

The second stage was to implement the second approach of the proposed algorithm. Therefore, the previous stage passes the deployed applications and hosting zones to this stage. The point of this stage is to reduce the network traffic between the zones by redeploying each application into the zones based on its dependencies. The first step in this stage was to build a graph of vertices and edges that represents the application dependencies. \textit{DepTree} class was responsible for building the graph, and it represents the dependencies between applications as  edges. 

The second step, called \textit{maxcut} package, redeployed the applications to the zones by applying the Local Search algorithm. Local Search algorithm finds the maximum cut of the dependencies between applications that are in different zones. The function  \textit{MaxCut()} in the \textit{maxcut} package is  responsible for reducing those dependencies. Finally, this stage invoked the calculation package to calculate the traffic cost, load balancing, and the performance of the cloud container after reducing the dependencies.

The last step was to apply the last proposed approach of the algorithm. This step invoked the \textit{enhloadbalancing} package to maintain the load balancing of zones. It was believed that the second step unbalanced the zone loads. Therefore, it was aimed in this part of the algorithm to redeploy independent applications that were in overload zones to underloaded zones. This step enhances the load balancing of zones and enhances the cloud container performance. 

\chapter{Results}
This chapter focuses on the experiment's results.  The cases that were used to  evaluate the algorithm were one small and another large scenario. Table  6.1 shows the test case details.
\begin{center}
 \begin{tabular}{||c c c c ||} 
 \hline
 Test Case Number & Application Number & Zones Number & Dependencies Number   \\ [0.5ex] 
 \hline\hline
 1 & 40 & 4 & 22 \\ 
 \hline
 2 & 200 & 10 & 423 \\[1ex]
 \hline  

\end{tabular}
\\ Table 6.1: Test Cases Table 
\end{center}
 The small test case has lower dependencies between applications, but the second test case has heavy dependencies between applications. The dependencies between applications in the second test case show that most of the applications depended on other applications.  It was considered that all applications arrived at the same time to the containers.  This section is divided into two sections based on test cases.  
\section{First Test Case}
\subsection{First Approach "Load Balancing of containers" }
 Firstly, the simulator initialised the information of the applications and zone configurations. Secondly, the round-robin algorithm scheduler used the initialisation information to deploy the applications into the zones. The scheduler deployed the applications into the zones equally. Each zone had ten applications based on the application requirements and zone capacities. Figure 6.1.1 shows the load balance of the applications in the zones. 

\begin{figure}
\centering
zone0   :++++++++++

zone1   :++++++++++

zone2   :++++++++++

zone3   :++++++++++

Figure 6.1.1: The result of first test case in first approach 
\end{figure}

The simulator shows that each zone has applications that fit in the zone resources. Table 6.1.2 shows the names of applications and the names of their zone hosts.
\begin{center}
 \begin{tabular}{|| c c ||} 
 \hline
 zone0 &[A2 A3 A7 A8 A10 A11 A12 A13 A14 A16]\\
 \hline
 zone1 &[A0 A1 A5 A17 A19 A20 A21 A25 A26 A27]\\
\hline
 zone2& [A4 A23 A28 A29 A30 A31 A32 A33 A34 A35]\\
 \hline
 zone3 &[A6 A9 A15 A18 A22 A24 A36 A37 A38 A39]\\
 [1ex]
 \hline  
\end{tabular}
\\Table 6.1.2: Applications in each zone
\end{center}
This approach does not consider the dependencies between applications while deploying applications. For example, the matrix in figure 8.1 in the Appendix shows the dependencies between the applications. The matrix considers zero as no dependency between the application in a column and the application in a row. On the other hand, 1 indicates that there is a dependency between the application in the column and the application in the row. The matrix shows that there is a dependency between application A0 and A14; however, those two applications are located in different zones, as shown in the table 6.1.2. 

The simulation showed that load balancing is optimum with 100\%. The simulator calculated the coefficient variation (CV) for the zones, and it was zero, which means that each zone hosts the same number of applications. However, the dependent applications were deployed onto different zones and that caused network traffic. The simulation also showed the network traffic cost to be 3820 for each application that used the network traffic ones. The simulator calculated the proportion of the network traffic cost for all applications using network traffic once each and it was 3.41 \% (Table 6.1.3).  However, it was assumed that applications continued using the network traffic for a period of time. Also, the simulator allowed each application to request or respond ten times. Table 6.1.3 shows that the cost of using network traffic increased and the proportion of using the bandwidth raised to 34.10 \%.  The accumulated network traffic cost reduced the overall cloud container performance.  Figures 6.1.4 and 6.1.5 show the proportion of load balance, network traffic, and  cloud container performance. 

 \begin{tabular}{| l| c |}
 \hline
 \textbf{Description} & \textbf{Value}\\
 \hline
 Number of Dependencies Between Application in Different zones & 15\\
 \hline
 Number of Dependencies Between Application in the Same zones & 7\\
 \hline
The Total Traffic Cost&  3820 \\
\hline

The Total Dependencies Between Applications&  22 \\
\hline
 Edges Between Application in The Same Zone & 82 edges\\
 \hline
 Edges Between Application in different Zones & 30 edges\\
 \hline
 Moving Application Cost& 0 \\
 \hline
Total proportion Traffic Used& 3.4107142857142856 \\
\hline
Total Proportion traffic Used 10X& 34.107142857142854 \\
\hline
Coefficient Variation &  0\\
\hline
Load balancing&  100\\
\hline

 \hline
 \end{tabular}
 
 \begin{center}
 Table 6.1.3: The First test case result 
 \end{center}
\begin{center}
\begin{bchart}[step=10,max=100]
        \bcbar[text=load balance]{100} 
            \smallskip
        \bcbar[text=network traffic ]{34.10} 
            \medskip
            
    \end{bchart}
  \\ Figure 6.1.4: Load balancing and network traffic proportion  
\end{center}

\begin{center}
\begin{tikzpicture}
\begin{axis}[
    title={},
    xlabel={Time},
    ylabel={Proportion},
    xmin=0, xmax=11,
    ymin=0, ymax=160,
    xtick={0,1,2,3,4,5,6,7,8,9,10,11},
    ytick={0,20,40,60,80,100,120,140,160},
    legend pos=north west,
    ymajorgrids=true,
    grid style=dashed,
]

\addplot[
    color=blue,
    mark=square,
    ]
    coordinates {(0,100)(1,100)(2,100)(3,100)(5,100)(6,100)(7,100)(8,100)(9,100)(10,100)
    
    };
  
    \addplot[
    color=orange,
    mark=square,
    ]
    coordinates {(0,0)(.11,3.4)(2,6.82)(3,10.23)(5,13.64)(6,17.05)(7,20.46)(8,23.87)(9,27.28)(10,34.1)
    
    };
    \legend{load balancing,  network traffic cost }
\end{axis}

\end{tikzpicture}\\
Figure 6.1.5: Network traffic increasing rate during experiment
\end{center}

\subsection{Second Approach ”Reducing the Network Traffic Cost”}
In this section, an explanation is given on the ways the  algorithm reduces the network traffic cost between applications that are in different zones. This situation is handled by solving the maximum cut problem. The simulator used the local search algorithm to solve this problem. The experiment shows that the total of the dependencies between the applications was 22. The Fat-Tree is the topology of the container network architecture and the simulator calculated the overall edges that connected all dependent applications. The simulator searched for all paths between source containers and destination containers, but the experiment considered just the shortest paths.  Table 6.1.3 shows the network traffic details. 

However, this methodology for reducing the traffic cost is designed to cut the maximum dependencies between applications in different zones. A gradual reduction of this kind of dependencies was one of the aims.  The dependencies were gradually reduced by as much as the following: 20\%,40\%,60\%, and 100 \%.  In each cut, the number and the name of applications that had been moved are shown. Moreover, the simulator shows the output of  load balancing and traffic cost for each cut. By using this methodology, it is possible to build a matrix that gives an indication of the container system performance, based on load balancing and network traffic. The following figures show the algorithm output of reducing the traffic cost by reducing the dependencies between applications in different zones. 

\begin{figure}[!h]
  \begin{minipage}[b]{0.35\textwidth}
    \includegraphics[width=1.2\textwidth]{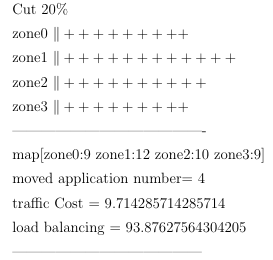}
    Figure 6.2.1:reduce 20\% of application Dependencies.
  \end{minipage}
  \begin{minipage}[b]{0.35\textwidth}
    \includegraphics[width=1.2\textwidth]{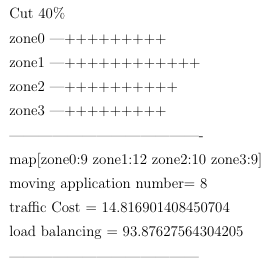}
    Figure 6.2.2:reduce 40\% of application Dependencies.
  \end{minipage}
  \end{figure}

\begin{figure}[!h]
  \begin{minipage}[b]{0.35\textwidth}
    \includegraphics[width=1.2\textwidth]{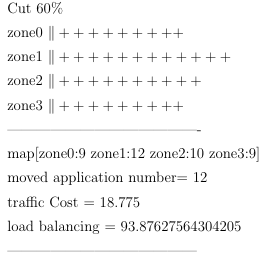}
    Figure 6.2.3: Reduce 60\% of application Dependencies.
  \end{minipage}
  \begin{minipage}[b]{0.35\textwidth}
    \includegraphics[width=1.2\textwidth]{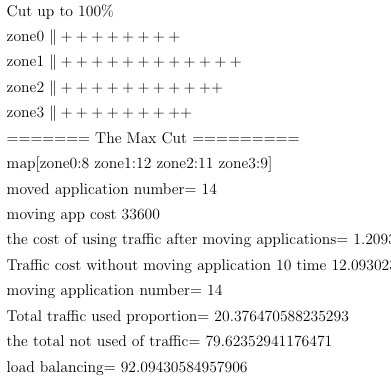}
   Figure 6.2.4: Reduce 100\% of application Dependencies.
  \end{minipage}
  \end{figure}
  
The most important feature of this algorithm is that it gradually reduces the dependencies between the applications in different zones by up to 93\%.  First, the algorithm reduced the application dependencies by 20\% by moving four applications to their dependent applications, as can be seen in Figure 6.2.1. This step reduced the total traffic cost by 9.714\%. The second step moved four applications and reduced the dependencies by 40\%. Moreover, this cut increased the reduction of traffic cost up to 14.81 per cent (Figure 6.2.2). Thirdly, the algorithm increased the reduction of traffic cost by up to 18.77\% by moving four applications from their zones to other zones (Figure 6.2.3).  In the last step, the algorithm reduced the network traffic in total from 34.10\% to 12.01\% by moving the last two applications to their new zones. The algorithm reduced 14 dependencies between applications that are in different zones of 15 dependencies, and it grouped the dependent applications into the same zones. Moving applications between zones increased the traffic cost for one time by up to 40\% of the network traffic bandwidth (Figure 6.2.4). 

On the other hand, the  algorithm reduced almost 64\% of the usage of the network traffic by applications without considering moving applications cost. Although moving applications between zones  had reduced the network traffic, it also reduced the load balancing among the containers  8\% less than the first approach. Table 6.2.1 shows the details of the main feature of this algorithm. Also, Figure 6.2.5 shows that the network traffic started with 50\% of network traffic bandwidth because there were 4 applications moved and the rest of the applications that were not moved kept using the network. Moving applications cost 40\% but the rest of the cost was because dependent applications were using the network. The usage of network traffic kept increasing until the third cut and started decreasing. When the dependent applications were moved the network traffic dropped to 1.2\%.
\begin{table}[!h]
  \begin{center}
    Table  6.2.1.Algorithm Result of Maximum Cut of Dependencies
    \label{tab:table1}
    \begin{tabular}{l|c} 
      \textbf{Equation Description 1} & \textbf{Value }\\
      \hline
     Map[zone0:8 zone1:12 zone2:11 zone3:9]
moving application number& 14\\
In The Same Zone Dependencies Before Max-Cut& 84 Edges \\
Outside Zones Dependencies Before Max-Cut &30Edges \\
In Depend After Max-Cut &84 Edges \\
Out Depend After Max-Cut &2 Edges \\
Number of Applications moved &14 Edges\\
Moving Applications cost &33600\\

 Dependency between application in same zone & 84 \\
 Dependency Between Applications In Different After  Max-Cut& 2\\

The cost of using traffic after moving applications &  1.2093023255813953\\
Traffic Cost Without Moving Applications 10 Time  &  12.093023255813954\\
Moving Applications Number &  14\\
Traffic for Moving Applications &  40\%\\
Total Traffic Cost  Proportion &  20.376470588235293\\
Total not Used of Traffic &  79.62352941176471\\
Load Balancing &   92.09430584957906\\

    \end{tabular}
  \end{center}
\end{table}
\begin{center}
\begin{tikzpicture}
  \begin{axis}[
        ybar, axis on top,
        title={Figure 6.2.5: Cut the Dependencies Gradualy },
        height=8cm, width=16cm,
        bar width=0.4cm,
        symbolic x coords={F 1, F 2, F 3, F 4, F 5,F 6, F 7, F 8, F 9},
        ymajorgrids, tick align=inside,
        major grid style={draw=white},
        enlarge y limits={value=.1,upper},
        ymin=0, ymax=100,
        axis x line*=bottom,
        axis y line*=right,
        y axis line style={opacity=0},
        tickwidth=0pt,
        enlarge x limits=true,
        legend style={
            at={(0.5,-0.2)},
            anchor=north,
            legend columns=-1,
            /tikz/every even column/.append style={column sep=0.5cm}
        },
        ylabel={Percentage (\%)},
        symbolic x coords={
           CUT20,
           CUT40,
           CUT60,
         CUT100,
         Time2,
         Time4,
         Time6,
         Time8,
         Time10
           },
       xtick=data,
       nodes near coords={
        \pgfmathprintnumber[precision=0]{\pgfplotspointmeta}
       }
    ]
    \addplot [draw=none, fill=blue!30] coordinates {
      (CUT20,4)
      (CUT40, 4) 
      (CUT60,4)
      (CUT100,2) 
      (Time2,0) 
      (Time4,0) 
      (Time6,0) 
      (Time8,0) 
      (Time10,0)
       };
   \addplot [draw=none,fill=red!30] coordinates {
      (CUT20,93.87)
      (CUT40, 93.87) 
      (CUT60,93.87)
      (CUT100,92.09)
      (Time2,92.09) 
      (Time4,92.09) 
      (Time6,92.09) 
      (Time8,92.09) 
      (Time10,92.09)
       };
   \addplot [draw=none, fill=green!30] coordinates {
      (CUT20,40)
      (CUT40, 40) 
      (CUT60,40)
      (CUT100,40)
      (Time2,0) 
      (Time4,0) 
      (Time6,0) 
      (Time8,0) 
      (Time10,0)
       };
     
\addplot[draw=blue,ultra thick,smooth] 
    coordinates {(F 1,49.9)(F 2,54.8)(F 3,58.7)(F 4,52)(F 5,1.2)(F 6,3.6)(F 7,6)(F 8,8.4)(F 9,12)};
    \legend{Num of Moved App ,Load Balancing,Moving app Traffic Cost,Total Traffic Cost}
  \end{axis}
  \end{tikzpicture}
 
\end{center}

 It confirmed that the algorithm increased the network traffic by up to 40\% during the moving application, but that was for one time. Redeploying applications caused reducing the load balancing of containers from 100\% to 92\%.
 
\subsection{Enhancing The Load Balancing Approach}
In this section, the aim to enhance the load balancing by redeploying independent applications from overloaded zones to under-loaded zones.  Applications are gradually redeployed. This is followed by monitoring the cost of moving applications and load balancing because dependent applications will not be redeployed, thus the total network cost will not be affected. This algorithm considers the  requirements of applications and  capacities of zones while redeploying  applications from overloaded zones. 

The simulation started by redeploying independent applications from overloaded zones to under-loaded zones, and this step costs 40\% of the network traffic because of the size of the applications.  The simulator redeployed the applications gradually. Figures 6.3.1 and 6.3.2 show the number of moved applications and the load balancing during the moving process. The algorithm redeployed one application from the overloaded zone to under-loaded zones and that increased the load balancing to 93.87. The second time, the algorithm moved three applications from the overloaded zones to under-loaded zones and that  enhanced the load balancing for all zones to 100\%. 

\begin{figure}[!h]
  \begin{minipage}[b]{0.5\textwidth}

zone0  :++++++++\\
zone1  :+++++++++++\\
zone2  :+++++++++++\\
zone3  :++++++++++\\
map[zone0:8 zone1:11 zone2:11 zone3:10]\\ 
number \ of \ moving \ application = 1\\
load \ balancing = 93.87627564304205\\
  Figure 6.3.1: The load balancing the first attempt 
  \end{minipage}
  \begin{minipage}[b]{0.5\textwidth}

zone0  :++++++++++\\
zone1  :++++++++++\\
zone2  :++++++++++\\
zone3  :++++++++++\\
map[zone0:10 zone1:10 zone2:10 zone3:10] \\
number \ of \ moving \ application = 3\\
load\ balancing = 100\\
      Figure 6.3.2: The load balancing the second attempt 
  \end{minipage}
  \end{figure}
The figure 6.3.3 shows the overall values of the algorithm including the processing of the reduced dependencies, network traffic cost, and enhanced the load balancing. 
\begin{center}
\begin{tikzpicture}
  \begin{axis}[
        ybar, axis on top,
        title={Figure 6.3.3: Overall values of the algorithm processing },
        height=8cm, width=16cm,
        bar width=0.4cm,
        symbolic x coords={F 1, F 2, F 3},
        ymajorgrids, tick align=inside,
        major grid style={draw=white},
        enlarge y limits={value=.1,upper},
        ymin=0, ymax=100,
        axis x line*=bottom,
        axis y line*=right,
        y axis line style={opacity=0},
        tickwidth=0pt,
        enlarge x limits=true,
        legend style={
            at={(0.5,-0.2)},
            anchor=north,
            legend columns=-1,
            /tikz/every even column/.append style={column sep=0.5cm}
        },
        ylabel={Percentage (\%)},
        symbolic x coords={
           1St-Approach,2Nd-Approach,3Rd-Approach
           },
       xtick=data,
       nodes near coords={
        \pgfmathprintnumber[precision=0]{\pgfplotspointmeta}
       }
    ]
    \addplot [draw=none, fill=blue!30] coordinates {
      (1St-Approach,100)
      (2Nd-Approach,92)
      (3Rd-Approach,100)
       };
   \addplot [draw=none,fill=red!30] coordinates {
      (1St-Approach,15)
      (2Nd-Approach,1)
      (3Rd-Approach,1)
       };
   \addplot [draw=none, fill=green!30] coordinates {
      (1St-Approach,0)
      (2Nd-Approach,40)
      (3Rd-Approach,20)
       };
     
\addplot[draw=blue,ultra thick,smooth] 
    coordinates {(F 1,34.10)(F 2,12.01)(F 3,12.01)};
    \legend{load balancing ,\# of dependencies,Moving app Traffic Cost,Total Traffic Cost}
    \addplot[draw=red,ultra thick,smooth] 
    coordinates {(F 1,100)(F 2,92)(F 3,100)};
    
  \end{axis}
 
  \end{tikzpicture}
  
\end{center}
 
\section{Second Test Case}
This test case has a high density of dependencies between applications. Also, it is a large scale scenario of 200 applications. The 200 applications were deployed into ten zones.  With this high density of dependencies between applications, it was expected  that the algorithm would enhance overall container system performance.   
\subsection{First Approach ”Container Load Balancing”} 

The simulation ran the second test case, and it deployed the applications on the zones fairly with CV= 0. Figure  6.4.1 shows the deployment results. The result of this data changed slightly because of the number of applications' dependencies. The figure  shows that the load balancing was 100\% during the first approach test period.

\begin{figure}[h]
  \begin{minipage}[b]{0.9\textwidth}
zone0   :++++++++++++++++++++\\
zone1   :++++++++++++++++++++\\
zone2   :++++++++++++++++++++\\
zone3   ;++++++++++++++++++++\\
zone4   :++++++++++++++++++++\\
zone5   :++++++++++++++++++++\\
zone6   :++++++++++++++++++++\\
zone7   :++++++++++++++++++++\\
zone8   ;++++++++++++++++++++\\
zone9   :++++++++++++++++++++\\
\end{minipage}
\begin{minipage}[b]{0.9\textwidth}

 The Number of Dependencies between application In The Same Zones=33\\
 The Number of Dependencies between application In The Different Zones=281\\
The Total Traffic Cost=  68760 \\
 The Number of Edges Between Application in Same Zone = 1256 \\
 The Number of Edges between application In The Different Zones = 562 \\
The Total Proportion Traffic Cost= 3.782178217821782 \\
The Total Proportion Traffic Cost 10X Time= 37.821782178217816 \\
The Load Balancing=  100\\
CV= 0 
\\
\begin{center}
    Figure 6.4.1: Second Test Case Output
  \end{center}
    \end{minipage}
    \end{figure}
  
The figure 6.4.2 shows the traffic consumption by the applications that have dependencies. Also, deployed applications had 281 dependencies between zones and 33 dependencies between applications in the same zones. The figure shows that the network traffic had gradually increased by 3.7\%  during the time of the experiments, while load balancing remained unchanged during experiment time. 
\begin{center}
\begin{tikzpicture}
  \begin{axis}[
        ybar, axis on top,
        title={First Approach Load Balancing \& Traffic Cost Proportion },
        height=8cm, width=16cm,
        bar width=0.4cm,
        symbolic x coords={F 1, F 2, F 3,F 4, F 5, F 6,F 7, F 8, F 9,F 10},
        ymajorgrids, tick align=inside,
        major grid style={draw=white},
        enlarge y limits={value=.1,upper},
        ymin=0, ymax=100,
        axis x line*=bottom,
        axis y line*=right,
        y axis line style={opacity=0},
        tickwidth=0pt,
        enlarge x limits=true,
        legend style={
            at={(0.5,-0.2)},
            anchor=north,
            legend columns=-1,
            /tikz/every even column/.append style={column sep=0.5cm}
        },
        ylabel={Percentage (\%)},
        symbolic x coords={
           1,2,3,4,5,6,7,8,9,10
           },
       xtick=data,
       nodes near coords={
        \pgfmathprintnumber[precision=0]{\pgfplotspointmeta}
       }
    ]
    \addplot [draw=none, fill=yellow!30] coordinates {
      (1,100)
      (2,100)
      (3,100)
      (4,100)
      (5,100)
      (6,100)
      (7,100)
      (8,100)
      (9,100)
      (10,100)
       };
   \addplot [draw=none,fill=blue!30] coordinates {
      (1,3.7)
      (2,7.4)
      (3,11.1)
      (4,14.8)
      (5,18.5)
      (6,22.2)
      (7,25.90)
      (8,29.6)
      (9,33.3)
      (10,37)
       };

\addplot[draw=blue,ultra thick,smooth] 
    coordinates {(F 1,)
      (F 2,)
      (F 3,)
      (F 4,)
      (F 5,)
      (F 6,)
      (F 7,)
      (F 8,)
      (F 9,)
      (F 10,)};
    \legend{load balancing ,Total Traffic Cost}
   
  \end{axis}
 
  \end{tikzpicture}
\\Figure 6.4.2
\end{center}

\subsection{Second Approach ”Reducing the Network Traffic Cost”}

 The dependencies were sequentially reduced by 20\%, 40\%,60\%,80\%,and 100\% of the application dependencies. The first cut of dependencies reduced the network traffic by 9.93\% of the total network traffic, which was 37.8. Moreover, it moved 62 applications to new zones with their neighbours. The most significant part of this step was that the load balancing had not changed and remained 100\%. The second cut reduced the network traffic, by 4.9\% of the total network traffic by moving 63 applications between zones. In this step, the load balancing was reduced to 95.5\%. The third cut of dependencies moved 63 applications to new zones and reduced the network traffic by 3.78\%. The fourth cut reduced network traffic by up to 21.20\% by moving 63 applications to their neighbours' zones. The last cut reduced the overall network traffic cost from 37.8\% to 22.09\%. Overall, this algorithm reduced the network traffic cost caused by dependencies between applications by up to 69.5\%. The total number of moved applications was 278.

However, the load balancing dropped gradually from 100\% to 86.07\% because of moving applications between zones. The Figures from 8.2 to 8.6 in the appendix  show the progress of the five steps of cutting the dependencies between applications. Moreover, Figure 6.4.4 shows the load balancing and network traffic cost during the reductions of dependencies. The network traffic was 38\% at the beginning of reducing the dependencies and started increasing because of moving applications. The maximum network traffic was 50\%  after finishing moving applications the network traffic dropped to 2.6\%, then because the network was consumed by dependent applications that are still in different zones the network traffic kept growing up to 26\% at the end of the experiment.

\begin{center}
\begin{tikzpicture}
  \begin{axis}[
        ybar, axis on top,
        title={Cut the Dependencies Gradualy },
        height=8cm, width=15cm,
        bar width=0.4cm,
        symbolic x coords={F 1, F 2, F 3, F 4, F 5,F 6, F 7, F 8, F 9,F 10},
        ymajorgrids, tick align=inside,
        major grid style={draw=white},
        enlarge y limits={value=.3,upper},
        ymin=0, ymax=100,
        axis x line*=bottom,
        axis y line*=left,
        y axis line style={opacity=2},
        tickwidth=0pt,
        enlarge x limits=true,
        legend style={
            at={(0.5,-0.4)},
            anchor=north,
            legend columns=-1,
            /tikz/every even column/.append style={column sep=0.4cm}
        },
        ylabel={Percentage (\%)},
        symbolic x coords={
         CUT20,
         CUT40,
         CUT60,
         CUT80,
         CUT100,
         Time2,
         Time4,
         Time6,
         Time8,
         Time10
           },
       xtick=data,
       nodes near coords={
        \pgfmathprintnumber[precision=0]{\pgfplotspointmeta}
       }
    ]
    \addplot [draw=none, fill=blue!30] coordinates {
      (CUT20,62)
      (CUT40, 63) 
      (CUT60,63)
      (CUT80,62)
      (CUT100,27) 
      
      (Time2,0) 
      
      (Time4,0) 
      
      (Time6,0) 
      
      (Time8,0) 
      
      (Time10,0)
       };
   \addplot [draw=none,fill=red!30] coordinates {
      (CUT20,100)
      (CUT40, 96.53) 
      (CUT60,87.45)
      (CUT80,85.21)
      (CUT100,86.01)
      
      (Time2,86.01) 
      
      (Time4,86.01) 
      
      (Time6,86.01) 
      
      (Time8,86.01) 
      
      (Time10,86.01)
       };
   \addplot [draw=none, fill=green!30] coordinates {
      (CUT20,40)
      (CUT40, 40) 
      (CUT60,40)
      (CUT80,40)
      (CUT100,40)
      
      (Time2,0) 
      
      (Time4,0) 
      
      (Time6,0) 
      
      (Time8,0) 
      
      (Time10,0)
       };
     
\addplot[draw=blue,ultra thick,smooth] 
    coordinates {(F 1,37.82)(F 2,49.93)(F 3,44.9)(F 4,43.76)(F 5,42.59)(F 6,40.08)(F 7,5.2)(F 8,10.4)(F 9,15.6)(F 10,26)};
    \legend{Num of Moved App ,Load Balancing,Moving app Traffic Cost,Total Traffic Cost}
  \end{axis}
  \end{tikzpicture}
  Figure 6.4.4
\end{center}

\subsection{Enhancing The Load Balancing Approach }
 
In the first phase of this experiment, the simulator gradually redeployed independent applications from overload zones to the less-loaded zones four times. The first redeployment involved moving 24 applications from overload zones to under-load zones and that increase the load balance from 86.01\% to 93.25\%. This stage enhanced the load balancing by 7.7\% figure 8.7 and figure 8.8 in appendix.

In the second phase, the simulator redeployed 13 applications between the zones that included independent applications and that were almost fully loaded. This step raised the load balancing to 100\%.  In conclusion, this algorithm increased the load balancing for both phases from 86.1\% to  100\% by almost 14.45\%. Figures 8.9 to 8.10 in the Appendix show the process steps of these two phases, and the zone changes during redeploying applications. Figure 6.4.5 shows the enhancement of the load balancing while redeploying the applications in this part.  

\begin{center}
\begin{tikzpicture}
  \begin{axis}[
        ybar, axis on top,
        title={Enhancing The Load Balancing },
        height=10cm, width=13cm,
        bar width=1.2cm,
        symbolic x coords={F 1, F 2, F 3, F 4, F 5,F 6},
        ymajorgrids, tick align=inside,
        major grid style={draw=white},
        enlarge y limits={value=.3,upper},
        ymin=0, ymax=100,
        axis x line*=bottom,
        axis y line*=left,
        y axis line style={opacity=2},
        tickwidth=0pt,
        enlarge x limits=.25,
        legend style={
            at={(0.5,-0.4)},
            anchor=north,
            legend columns=-1,
            /tikz/every even column/.append style={column sep=0.2cm}
        },
        ylabel={Percentage (\%)},
        symbolic x coords={
           1St Phase,2nd Phase},
       xtick=data,
       nodes near coords={
        \pgfmathprintnumber[precision=2]{\pgfplotspointmeta}
       }
    ]
\addplot [draw=none, fill=green!30] coordinates {(1St Phase, 86.1)(2nd Phase, 100) };
\addplot [draw=none, fill=blue!30]coordinates {(1St Phase, 93.25)(2nd Phase,100 )};  

\end{axis}  
\end{tikzpicture}  
\\Figure 6.4.5: Enhance Load balancing for The Second Test Case
\end{center}
\section{Research Significance and Findings}
\definecolor{Gray}{gray}{0.9}
\definecolor{LightCyan}{rgb}{0.88,1,1}
\begin{table}
\centering
\begin{tabular}{|l|l|l|l|l|l|l|l|l|l|l|l|l|l|} 
\hline
\rowcolor{Gray}
 \rotatebox{90}{Test Case}  & \rotatebox{90}{Approach} &  \rotatebox{90}{\# Application}& \rotatebox{90} {\# Zones}&\rotatebox{90}{Bandwidth Gbps}& \rotatebox{90}{Network Throughput} &\rotatebox{90}{Traffic Cost \%} &\rotatebox{90}{CV}  & \rotatebox{90}{Load Balancing\%}&\rotatebox{90}{\# Dependencies Cut (Maximum cut \%)} & \rotatebox{90}{Decrease Network Traffic \%+Moving app \%}& \rotatebox{90}{Decrease Network Traffic}  &\rotatebox{90}{\#  Applications redeployed} &\rotatebox{90}{Enhance Load Balancing \%} \\ 

\hline

\multirow{10}{*}{\rotatebox{90} {First Test Case}} & 1St & 40 &4  & 1 & 3840 & 33.68 & 0 & 100 & 0&0 &33.68  & 0  &0  \\ 
 \cline{2-14}
 
 &\multicolumn{5}{l}{\cellcolor{yellow}Total For First Approach } & \cellcolor{green}33.68 & 0 & 100 &0& 0 & \cellcolor{green}33.68 & 0 & 0  \\

\cline{2-13}
 &  & 40 &4  &1  & 9600 & 40 & 1.22 &93.87  & 4& 9.71& 12.09 & 0  &-6.13  \\ 
 \cline{3-14}
 & & 40 &4 & 1 & 48000 & 40 & 1.22 & 93.87 & 4&5.1 & 12.09 &  0 & 0\\ 
\cline{3-14}
 & 2nd & 40 & 4 & 1 & 160000 & 40 & 1.22 & 93.87 & 4 &3.96& 12.09 & 0  &0\\ 
\cline{3-14}
 &  & 40 & 4 & 1 & 33600 &40  & 1.5 & 92.09 &  2& 1.6&12.09 &0  &-1.78\\ 
 \cline{2-14}
 &\multicolumn{6}{l}{\cellcolor{yellow}Total For Second Approach}& 1.5&\cellcolor{red}92.09 &\cellcolor{LightCyan}14 &\cellcolor{LightCyan}20.37&\cellcolor{LightCyan}12.09&0 & \cellcolor{red}-7.91\\
\cline{2-14}
 &  & 40 & 4 & 1 &2400  & 20.37 &1.22  & 93.87 & 0 &0& 0 & 1  &1.78\\ 
\cline{3-14}
 &  & 40 & 4 & 1 & 7200 & 20.37 & 0 & 100 & 0 & 0 &0& 3  &6.13\\ 

 \cline{2-14}
 &\multicolumn{5}{l}{\cellcolor{yellow}Total For Third Approach} & \cellcolor{LightCyan}20.37 & 0 & \cellcolor{orange}100&0 & 0&\cellcolor{LightCyan}12.09&\cellcolor{orange}13 &\cellcolor{orange}7.91\\
\hline
 \multirow{10}{*}{\rotatebox{90} {Second Test Case}} & 1st & 200 &10  & 1 & 68760 & 37.82 & 0 & 100 & 0 & 0 &37.82& 0  &0\\ 
 \cline{2-14}
 &\multicolumn{5}{l}{\cellcolor{yellow}Total For First Approach } & \cellcolor{green}37.82 & 0 & 100 &0 &  0 &\cellcolor{green}37.82& 0 & 0  \\
\cline{2-14}
 &  & 200 & 10 & 1 & 148800 & 40 & 0 & 100 & 62 & 9.93 &26.31& 0  &0\\ 
 \cline{3-14}
 &  & 200 & 10 & 1 & 151200 & 40 & 2.19 & 96.53 & 63 & 4.9 &26.31& 0  &-3.4\\ 
\cline{3-14}
 & 2nd &200  &  10& 1 & 300000 & 40 & 7.93 & 87.45 & 63 & 3.78 & 26.31 &0 &-9.08\\ 
\cline{3-14}
 &  & 200 & 10 & 1 & 151200 & 40 & 9.34 & 85.21 & 63 & 2.59 & 26.31& 0 &-2.24\\ 
\cline{3-14}
 &  & 200 &  10& 1 & 46800 & 40 & 8.484 & 86.01 & 27 & 0.8 & 26.31& 0 &1\\ \cline{2-13}
 &\multicolumn{6}{l}{\cellcolor{yellow}Total For Second Approach}& 8.484&86.01 &\cellcolor{LightCyan}278 &\cellcolor{LightCyan}22.09&\cellcolor{LightCyan}26.31 &0&\cellcolor{red} -13.72\\
\cline{2-14}
 &  & 200 & 10 & 1 & 28800 & 22.09 &7.56  &86.01  & 0 & 0& 0& 0  &0\\ 
\cline{3-14}
 & 3rd & 200 & 10 & 1 & 48000 & 22.09 & 6.29 & 93.25 & 0 & 0 &0& 24  &7.7\\ 
\cline{3-14}
 &  & 200 & 10 & 1 & 48000 &  22.09& 6.29 &  100& 0 & 0 &0& 13  &6.75\\ 
\cline{3-14}
 &  & 200 & 10 & 1 & 48000 &  22.09&  6.29&  100& 0 &  0& 0&0 &0\\ 

\cline{2-14}
 &\multicolumn{5}{l}{\cellcolor{yellow}Total For Third Approach} & \cellcolor{LightCyan}22.09 & 0 & \cellcolor{orange}100&0 &0 &0&\cellcolor{orange}37 &\cellcolor{orange}14.45\\
\hline

\end{tabular}
\\Figure 6.5.1
\end{table}

The matrix in Table 6.5.1 shows all the experiment outputs. The matrix is divided into two main parts which are the first test case and second test case. Each part is divided into three approaches: first approach, second approach and third approach. Each approach has a total that is highlighted with yellow colour. The significant findings are coloured to illustrate the result of the algorithm. The indication of the colours is the following:
\begin{enumerate}
    \item Green colour indicates the network traffic cost proportion for the first approach. 
\item Light blue colour indicates the dependencies that were cut by the maximum cut (local Search algorithm) and the proportion of the reduction in the network traffic cost which is the second approach. 
 \item Red colour indicates the proportion of load balancing changes due to the movement of the applications.
  \item Orange colour indicates the enhanced load balancing and the number of independent applications that were redeployed to balance the container loads which is the last approach. 
\end{enumerate}

\begin{center}
\begin{figure}[h!]
 \scalebox{0.7}{
    \includegraphics{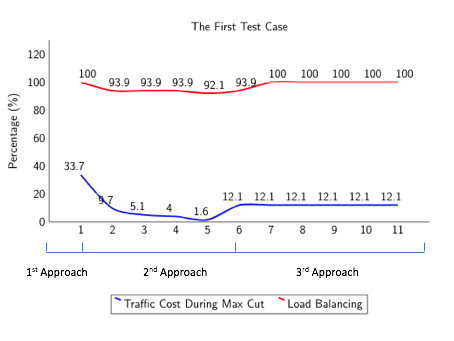}}
    \\ Figure 6.5.2: All Approaches Result for first Test Case
   \end{figure}

   \begin{figure}[h!]
    \scalebox{0.7}{
    \includegraphics{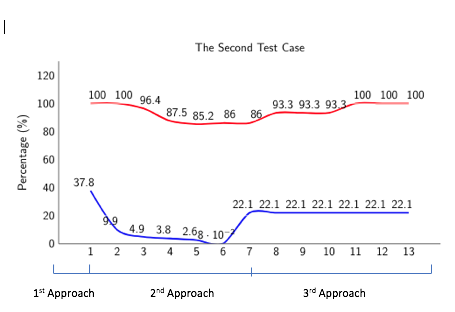}
 }
\\ Figure 6.5.3: All Approaches Result for Second Test Case
\end{figure}
\end{center}

Figures(6.5.2 and 6.5.3) show the overall results of the algorithm. The behaviour of the algorithm was similar for both test cases. Firstly, the algorithm distributed the application equally into zones without considering the network traffic for both test cases. During this stage, the load balancing for all zones was 100\%.

When the algorithm started the second approach, which reduced dependencies between applications that were in different zones, the cost began to drop sharply.  For example, the cost of traffic in the first test case dropped from 33.7 to 1.2. In the second test case , the cost of network traffic dropped from 37.8 to 2.209. However, the cost of network traffic started increasing by the end of this approach because of the dependencies between applications that had not been moved by the approach. The cost of all applications that need network traffic if each application uses the network traffic ones is 1.2 in the first test case  and 2.209 for the second test case. Still, the number increased because the applications continued use of the network traffic for t time.

One of the most important features of the algorithm is that, the algorithm was moving applications between zones gradually during reducing dependencies between applications and during enhancing the load balancing. This approach of moving applications reduces the consumption the network traffic efficiently. Also, it reduces the network congestion and as a result, it reduces the probability of losing packets. 

On the other hand, during the second approach, the load balance in the two test cases had been reduced and fluctuated because of moving applications between the zones. However, by the end of the third approach, the first test case's and second test case's load balancing enhanced to 100\% by the last approach of the research algorithm.

\section{Adding Applications Scenario}
 It was assumed that all applications needing to be deployed into the containers would arrive at the same time. In this part of the research, the ability of the algorithm to deploy applications that would arrive after finishing deployment of current applications was tested. 
 
These steps tested the capability of the algorithm to efficiently distribute applications into containers by simulating a real container system and testing the behaviour of the algorithm. The simulation of the real container system  increased the dependencies between the newly arriving applications. It was expected that the algorithm would perform efficiently. However, because of the dependencies between the newly arriving applications and the deployed applications, there was no guarantee that the results would be the same as the previous results. 

In this test case, focus is on the dependencies between applications rather than on the number of arriving applications. Therefore, the number of arriving applications will be 20 applications for both test cases. The first test case will have at least one dependency between arriving applications. The second test case will have double dependencies between applications of the first test case. The following table shows the most important features of the results.

\subsection{Test Case Results}
The table 8.1 in the appendix  shows that each test case has 20 arrived applications to the containers system.  However, the fist test case had 16  dependencies between applications that are in different zones. The second test case had 40 dependencies between applications and 24 of these dependencies are between applications that are in different zones.  

On the other hand, due to the high density of the dependencies between applications that arrived and the applications that had been deployed on the zones' containers, the traffic cost increased to 24.4\% for the first test case that was 12.09 Figure 6.1.1 shows the arrived application result after the vertical black plot. Moreover, the traffic costs also increased form 22.1 \% to 27.44 after 20 applications arrived in the second test case. This increase was the result of the dependencies between applications. The simulator shows that the algorithm reduced the dependencies between applications in different zones from 16 to 4 in the first test case, and from 302 to 244 dependencies between applications in different zones. 

The algorithm also reduced the dependencies between application in the first test case and the second test case by 75\% and by 19\% respectively. The reason for the algorithm reducing 19\% of the dependencies between applications in different zones in the second test case is the high density of the dependencies between applications. However, the simulator shows that the algorithm reduces the traffic cost between applications in the first test case from 24.4 to 14.09 by 42\%.  In addition, the algorithm reduced the traffic cost in the second test case from 27.64 to 19.7 without considering the cost of moving applications. The algorithm reduced the traffic cost in the second test case after 20 applications arrived with 40 dependencies between applications by 28\% of the total throughput Figure 6.6.2 shows the arrived application result after the vertical black plot. 

The simulator shows that the load balancing was reduced from 100\% for both test cases because of moving applications and cut the dependencies between applications that were in different zones. However, the load balance was reduced  to 88.54\% and  to 89.81 in the first and the second test cases, respectively. However, the last approach of the algorithm enhanced the load balancing to 88.81 and 91.05 for the first test case and the second test case, respectively. Due to each of the arrived applications having at least one dependency with other applications, the last approach of the algorithm did not enhance the load balancing to 100\%. 
\begin{center}
\begin{tikzpicture}
 \begin{axis}[
        ybar, axis on top,
        title={The First Test Case With  20 newly arriving dependent applications  },
        height=8cm, width=15cm,
        bar width=0.4cm,
        symbolic x coords={F 1, F 2, F 3, F  4, F 5,F 6, F 7, F 8, F 9,F 10,F 11,F 12,F 13,F 14,F 15,F 16,F 17,F 18,F 19,F 20,F 21,F 22,F 23},
        ymajorgrids, tick align=inside,
        major grid style={draw=white},
        enlarge y limits={value=.3,upper},
        ymin=0, ymax=100,
        axis x line*=bottom,
        axis y line*=left,
        y axis line style={opacity=2},
        tickwidth=0pt,
        enlarge x limits=true,
       legend style={
            at={(0.5,-0.4)},
            anchor=north,
            legend columns=-1,
        /tikz/every even column/.append style={column sep=0.7cm}
       },
        ylabel={Percentage (\%)},
        symbolic x coords={
         1,2,3,4,5,6,7,8,9,10,11,12,13,14,15,16,17,18,19,20,21,22,23},
      xtick=data,
       nodes near coords={
        \pgfmathprintnumber[precision=0]{\pgfplotspointmeta}
       }
    ]
 
     \addplot[draw=blue,ultra thick,smooth] 
    coordinates {(F 1,33.7)(F 2,9.7)(F 3,5.1)(F 4,4)(F 5,1.6)(F 6,12.1)(F 7,12.1)(F 8,12.1)(F 9,12.1)(F 10,12.1)(F 11,12.1)(F 12,24.49)(F 13,14.32)(F 14,0.23)(F 15,1.4)(F 16,2.8)(F 17,4.2)(F 18,5.6)(F 19,7)(F 20,8.4)(F 21,9.8)(F 22,11.2)(F 23,14.0)};
    \addplot[draw=red,ultra thick,smooth] 
    coordinates {(F 1,100)(F 2,93.9)(F 3,93.9)(F 4,93.9)(F 5,92.1)(F 6,93.9)(F 7,100)(F 9,100)(F 11,100)(F 12,90.97)};
   
   \addplot[draw=orange,ultra thick,smooth] 
    coordinates{(12,90.97)(F 13,90.82)(F 14,90.13)(F 15,88.54)(F 16,88.09)(F 17,88.09)(F 18,88.09)(F 19,88.09)(F 20,88.81)(F 21,88.81)(F 22,88.81)(F 23,88.81)};
    \addplot[draw=black,ultra thick,smooth] 
     coordinates{(F 12,120)(F 12,1)};
    \legend{Traffic Cost ,Load Balancing,,Start Arrive New Application}
  \end{axis}
 
  \end{tikzpicture}
  Figure 6.6.1
\end{center}
\begin{center}
\begin{tikzpicture}
 \begin{axis}[
        ybar, axis on top,
        title={The Second Test Case With Arrive 20 Dependent Applications  },
        height=8cm, width=15cm,
        bar width=0.4cm,
        symbolic x coords={F 1, F 2, F 3, F  4, F 5,F 6, F 7, F 8, F 9,F 10,F 11,F 12,F 13,F 14,F 15,F 16,F 17,F 18,F 19,F 20,F 21,F 22},
        ymajorgrids, tick align=inside,
        major grid style={draw=white},
        enlarge y limits={value=.3,upper},
        ymin=0, ymax=100,
        axis x line*=bottom,
        axis y line*=left,
        y axis line style={opacity=2},
        tickwidth=0pt,
        enlarge x limits=true,
       legend style={
            at={(0.5,-0.4)},
            anchor=north,
            legend columns=-1,
        /tikz/every even column/.append style={column sep=0.7cm}
       },
        ylabel={Percentage (\%)},
        symbolic x coords={
         1,2,3,4,5,6,7,8,9,10,11,12,13,14,15,16,17,18,19,20,21,22},
      xtick=data,
       nodes near coords={
        \pgfmathprintnumber[precision=0]{\pgfplotspointmeta}
       }
    ]

\addplot[draw=blue,ultra thick,smooth] 
    coordinates {(F 1,37.82)(F 2,9.93)(F 3,4.9)(F 4,3.76)(F 5,2.59)(F 6,0.08)(F 7,22.09)(F 8,22.09)(F 9,22.09)(F 10,24.4)(F 11,17.52)(F 12,1.97)(F 13,3.94)(F 14,5.91)(F 15,7.88)(F 16,9.85)(F 18,13.79)(F 20,17.73)(F 22,19.7)};
    
    \addplot[draw=red,ultra thick,smooth] 
    coordinates {(F 1,100)(F 2,100)(F 3,96.35)(F 4,87.45)(F 5,85.21)(F 6,86.01)(F 7,86.01)(F 8,93.25)(F 9,100)};
    
   \addplot[draw=orange,ultra thick,smooth] 
    coordinates{(F 9,100)(F 10,92.72)(F 11,90.64)(F 12,89.91)(F 13,89.81)(F 14,90.17)(F 15,90.92)(F 16,91.05)(F 17,91.05)(F 18,91.05)(F 19,91.05)(F 20,91.05)(F 21,91.05)(F 22,91.05)};
    \addplot[draw=black,ultra thick,smooth] 
     coordinates{(F 9,120)(F 9,1)};
    \legend{Traffic Cost ,Load Balancing,,Start Arrive New Application}
  \end{axis}
  \end{tikzpicture}
 Figure 6.6.2
\end{center}


 

\section{Cloud Container Performance}

Figure 6.1 shows the container performance along with load balancing and network traffic: 
\begin{center}
\begin{figure}[h!]
 \scalebox{0.9}{
    \includegraphics{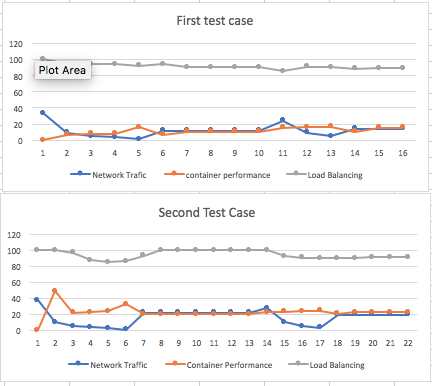}}
    \caption{}
    \end{figure}
\end{center}
In this section, the aim is to observe and monitor the performance of the containers. Figure 6.1 illustrates information related to performance during the running of the algorithm and its approaches. The figure also shows the performance when twenty applications arrived after the algorithm distributed applications into the zone containers. From the figure 6.1 it is evident that the algorithm increases the performance of the containers by reducing the network traffic. 

The figure demonstrates that the performance increases at points 2,4,5, and 6, while the algorithm reduces the network traffic even though the load balancing is reduced in the same period. Also, the figure shows wherever the network traffic increases the performance decreases and the other way around. Figure 6.1 shows that, in the first test case, the performance was at a peak when the algorithm reduced the network traffic to the lowest point at point 5.

Moreover, the figure shows that the performance of containers reached the maximum point when the algorithm sharply reduced the network traffic from 37.82\% to almost 10\% at the beginning of the maximum cut at point 2. It is true that load balancing affects the performance of the container, but it is not comparable with the network traffic because at point 6 for both test cases when the network traffic started increasing, the container's performance started decreasing even though the load balancing was enhanced by the algorithm.

\chapter{Conclusion}
Container technology has improved distributing applications on top of the host hardware and operating system. Container technology has also reshaped the virtualisation technology.  This new technology provides the users with a tremendous amount of hardware and software infrastructure. Container technology has been an environment for companies to offer significant services with high performance. However, due to the distributed applications into the containers sharing the host operating system and hardware, the contention of resources has been a major problem. Moreover because of the dependencies between applications in the cloud system the network traffic potentially causes latency, loses  packets, creates response delay, and delays the running of applications. Also, dependencies between applications increase the cost of network traffic. 

In this research, an algorithm that improves the performance of the could container system was proposed with the aim of reducing the cost of network traffic in container systems by reducing the dependencies between applications that are in different zones. It is believed that dependencies between applications in different zones increase the cost of network traffic more than the dependencies between applications that are in the same zone. Therefore, the aim was to group the applications that have dependencies in a small number of zones. The  algorithm was divided into three approaches, each of which had the purpose to achieve improvement in the performance of containers technology.

The algorithm was applied to two test cases: a small scale test case with 40 applications that were deployed into four zones with ten containers in each zone; the second test case with 200 applications that were deployed into ten zones with twenty containers in each zone. The first test case had normal dependencies between applications; however, the second test case had almost two dependencies for each application. The algorithm deployed applications into the zones equally for both test cases in the first approach. Moreover, by the end of the third approach, the algorithm reduced the dependencies between applications that were in different zones by 93\%, 69.5\% for both first and second test cases respectively, without affecting the load balancing of the applications. That improved the container performance by reducing the network traffic by 64\% and 50\% for both test cases respectively without affecting the load balancing. 

Moreover, Zhao et al. (2020)  consider each application has a dependency with only two neighbours and force each zone to have one application for minimising the problem of finding the path between applications. We consider each application can have many dependencies, and zones can hold more than one applications that have dependencies. Also we consider all probability of reducing network traffic and we maintain the load balance as well, in contrast of Zhao et al. (2020) who maintained the traffic cost by adjusting the load balance and network traffic.

Furthermore, the algorithm was designed to receive extra applications that arrived after distributing applications into the zones. The algorithm was investigated by simulating  real containers. It was assumed that the applications were possible to be deployed into the containers at any time. Therefore, twenty applications into each test case were added after the algorithm deployed the first group of applications. In this stage, the dependencies between arriving applications and already distributed applications were established with at least one dependency either with the deployed applications or with the new application. However, the  density of dependencies between applications in the second test case were high. The algorithm reduced the dependencies between applications that were in different zones by 75\% and 19\% in  first and second test cases, respectively. The reduction decreased the traffic cost by 42\% and 28\% for the first and second test cases, respectively. 

In conclusion, the algorithm shows the ability and flexibility to distribute applications into zone containers equally. Also, it demonstrates that it can receive and redeploy applications into zone containers at any time. The most important feature of the algorithm is that it is able to reduce the dependencies between applications in different zones and that it effectively reduces the cost of network traffic and boosts the performance of cloud container. 

\bibliography{bibliography}
\chapter{Appendix}
\section{Figures}
This section includes all figures that were not included in the main text.

\subsection{\textit{\textbf{The dependencies matrix figure:}}}

\begin{figure}[h!]
    \begin{center}
    \scalebox{0.90}{
    \includegraphics{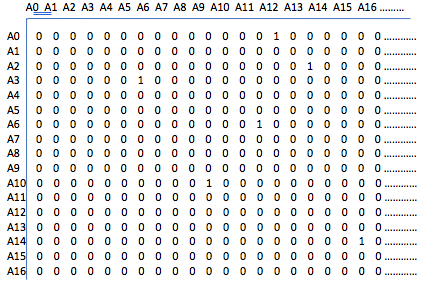}
    }
    \caption{Dependencies matrix\label{fig:FatTree}}
    \end{center}
\end{figure}  

\subsection{\textit{\textbf{The Maximum Cut of The Second Test Case}}}

This is the output of the maximum cut for the second test case \textbf{(Second Approach)}

\begin{figure}[h!]
zone0:++++++++++++++++++++\\
zone1:++++++++++++++++++++\\
zone2:++++++++++++++++++++\\
zone3:++++++++++++++++++++\\
zone4:++++++++++++++++++++\\
zone5:++++++++++++++++++++\\
zone6:++++++++++++++++++++\\
zone7:++++++++++++++++++++\\
zone8:++++++++++++++++++++\\
zone9:++++++++++++++++++++\\
map[zone0:20 zone1:20 zone2:20\\ 
zone3:20 zone4:20 zone5:20 \\zone6:20
zone7:20 zone8:20 zone9:20]\\
moving application number= 62\\
traffic Cost = 9.930097087378641\\
moving application number= 62\\
load balancing = 100 \\
\caption{Reducing the dependencies with 20\% }
\end{figure}
\begin{figure}[h!]
zone0:++++++++++++++++++++\\
zone1:+++++++++++++++++++++++\\
zone2:+++++++++++++++++++++++\\
zone3:+++++++++++++++++++++\\
zone4:++++++++++++++++++++\\
zone5:++++++++++++++++++++\\
zone6:++++++++++++++++++++\\
zone7:++++++++++++++++++++\\
zone8:+++++++++++++++\\
zone9:++++++++++++++++++\\
map[zone0:20 zone1:23 zone2:23\\ 
zone3:21 zone4:20 zone5:20 zone6:20 zone7:20 zone8:15 zone9:18]\\
moving application number= 125\\
traffic Cost = 14.83419689119171\\
Cost of Moving Application 40\\
load balancing = 96.53589838486225\\
\caption{Reducing the dependencies with 40\%}
\end{figure}
\begin{figure}[h!]
zone0:++++++++++++++++\\
zone1:+++++++++++++++++++\\
zone2:+++++++++++++++++++++\\
zone3:++++++++++++++++++++++++++++++++++++++++++\\
zone4:++++++++++++++++++++++\\
zone5:+++++++++++++++++++\\
zone6:+++++++++++++++++++\\
zone7:+++++++++++++++\\
zone8:+++++++++++\\
zone9:++++++++++++++++\\
map[zone0:16 zone1:19 zone2:21 zone3:42 zone4:22 zone5:19 zone6:19 zone7:15 zone8:11 zone9:16]\\
moving application number= 188\\
traffic Cost = 18.61467889908257\\
Cost of Moving Application 40\\
load balancing = 87.45009960198887\\ 
\caption{Reducing the dependencies with 60\%}
\end{figure}
\begin{figure}[h!]

zone0:++++++++++++\\
zone1:+++++++++++\\
zone2:++++++++++++++++++++++++++++\\
zone3:++++++++++++++++++++++++++++++++++++++++++\\
zone4:+++++++++++++++++++++++\\
zone5:+++++++++++++++++++++++\\
zone6:+++++++++++++++++++\\
zone7:++++++++++++++++++++\\
zone8:+++++++++++\\
zone9:+++++++++++\\
map[zone0:12 zone1:11 zone2:28 zone3:42 zone4:23 zone5:23 zone6:19 zone7:20 zone8:11 zone9:11]\\
moving application number= 251\\
traffic Cost = 21.202380952380953\\
Cost of Moving Application 40\\
load balancing = 85.21825450090552\\  
\caption{Reducing the dependencies with 80\%}
\end{figure}
\begin{figure}[h!]
zone0:++++++++\\
zone1:+++++++++\\
zone2:+++++++++++++++++++++++++++\\
zone3:+++++++++++++++++++++++++++++++++++++\\
zone4:++++++++++++++++++++++++\\
zone5:+++++++++++++++++++++\\
zone6:+++++++++++++++++++\\
zone7:++++++++++++++++++++++++++++\\
zone8:+++++++++++\\
zone9:++++++++++++++++\\
in Depn Before Maxcut 1256 \\
out Depn Before Maxcut 562 \\
in Depend After Maxcut 1256 \\
out Depend After Maxcut 278 \\
 dependency between application in same zone = 1256 \\
 dependency between application between zones = 278\\
the cost of using traffic after moving applications= 2.6310299869621905\\
Traffic cost without moving application 10 time  26.310299869621904\\
traffic for moving app= 40\\
Total traffic used  proportion= 22.09743910056215\\
total not used of traffic= 77.90256089943784\\
load balancing=  86.01786854589044\\
\caption{Reducing the dependencies with 100\%}
\end{figure}
\newpage

\textbf{The Load Balancing of The Second Test Case}
This is the output of the balancing the load for the second test case \textbf{(Third Approach)}
\begin{figure}[h!]

zone0  :++++++++\\
zone1  :+++++++++\\
zone2  :+++++++++++++++++++++++++++\\
zone3  :+++++++++++++++++++++++++++++++++++++\\\
zone4  :++++++++++++++++++++++++\\
zone5  :+++++++++++++++++++++\\
zone6  :+++++++++++++++++++\\
zone7  :++++++++++++++++++++++++++++\\
zone8  :+++++++++++\\
zone9  :++++++++++++++++\\
map[zone0:8 zone1:9 zone2:27 zone3:37 \\
zone4:24 zone5:21 zone6:19 zone7:28\\
zone8:11 zone9:16]\\
 dependency between application in same zone = 1256\\ 
 dependency between application between zones = 278\\
load balancing=  86.01786854589044\\
traffic= 22.09743910056215\\
\caption{Balancing the load first stage}
\end{figure}
\begin{figure}[h!]

zone0  :++++++++++++++++++++\\
zone1  :++++++++++++++++++++\\
zone2  :++++++++++++++++++++\\
zone3  :++++++++++++++++++++++++++\\
zone4  :++++++++++++++++++++\\
zone5  :++++++++++++++++++++\\
zone6  :++++++++++++++++++++\\
zone7  :+++++++++++++++++++++++++++\\
zone8  :+++++++++++\\
zone9  :++++++++++++++++\\
map[zone0:20 zone1:20 zone2:20\\
zone3:26 zone4:20 zone5:20 zone6:20 \\
zone7:27 zone8:11 zone9:16] \\
moving application number= 24\\
Cost of Moving Application 40\\
movAppCost 57600
load balancing = 93.25463121838398\\
\caption{Balancing the load first stage}
\end{figure}
\begin{figure}[h!]
zone0  :++++++++++++++++++++\\
zone1  :++++++++++++++++++++\\
zone2  :++++++++++++++++++++\\
zone3  :++++++++++++++++++++\\
zone4  :++++++++++++++++++++\\
zone5  :++++++++++++++++++++\\
zone6  :++++++++++++++++++++\\
zone7  :++++++++++++++++++++\\
zone8  :++++++++++++++++++++\\
zone9  :++++++++++++++++++++\\
map[zone0:20 zone1:20 zone2:20 zone3:20\\
zone4:20 zone5:20 zone6:20 zone7:20\\
zone8:20 zone9:20] \\
moving application number= 37\\
Cost of Moving Application 40\\
movAppCost 88800\\
load balancing = 100\\
\caption{Balancing the load second stage}
\end{figure}
\begin{figure}[h!]

 final result\\
zone0  :++++++++++++++++++++\\
zone1  :++++++++++++++++++++\\
zone2  :++++++++++++++++++++\\
zone3  :++++++++++++++++++++\\
zone4  :++++++++++++++++++++\\
zone5  :++++++++++++++++++++\\
zone6  :++++++++++++++++++++\\
zone7  :++++++++++++++++++++\\
zone8  :++++++++++++++++++++\\
zone9  :++++++++++++++++++++\\
map[zone0:20 zone1:20 zone2:20 \\
zone3:20 zone4:20 zone5:20 zone6:20\\
zone7:20 zone8:20 zone9:20] \\
moving application number= 37\\
Cost of Moving Application 40\\
movAppCost 88800\\
load balancing = 100\\
\caption{Balancing the load second stage}
\end{figure}

\newpage
\textbf{The Table Result for The Arrived Applications}

\begin{table}
\centering
\begin{tabular}{|l|l|l|l|l|l|l|l|l|l|l|l|l|l|} 
\hline
\rowcolor{Gray}
 \rotatebox{90}{Test Case}  & \rotatebox{90}{Approach} &  \rotatebox{90}{\# Deployed Application}& \rotatebox{90} {\# Arriveal Applications}&\rotatebox{90}{\# Zones}& \rotatebox{90}{\# Dependencies Between Zones} &\rotatebox{90}{Traffic Cost \%} &\rotatebox{90}{CV}  & \rotatebox{90}{Load Balancing\%}&\rotatebox{90}{\%Dependencies Cut between zones} & \rotatebox{90}{Decrease Network Traffic \%+Moving app \%}& \rotatebox{90}{Decrease Network Traffic}  &\rotatebox{90}{\#  Applications redeployed} &\rotatebox{90}{\%Enhance Load Balancing } \\ 

\hline

\multirow{6}{*}{\rotatebox{90} {1st test Case}} & 1St & 40 &20  & 4 &\cellcolor{green}16  &24.4  &0  &100  & 0& 0& 42.4 & 0  &  0\\ 
 \cline{2-14}
 
 &\multicolumn{5}{l}{\cellcolor{yellow}Total For First Approach } & \cellcolor{green}24.4&  0& 100 &0& 0 & 24.4& 0 &0   \\
 \cline{2-14}

\cline{2-14}
 &2nd  & 40 & 20 & 4 & \cellcolor{green}4&14.09 & 2.29 & 88.54 & 75 & 14.09 &42 &0&  0 \\ 
 \cline{2-14}
 &\multicolumn{5}{l}{\cellcolor{yellow}Total For Second Approach } & \cellcolor{green}14.09 & 2.29 &\cellcolor{orange} 88.54 &75& 14.09 & 42&0  & 0 \\
 \cline{2-14}

\cline{2-14}
 & 3rd & 40 & 20 & 4 & 4 & 14.09 &3.35  &88.81  & 0 & 14.09 &0& 1  &0.27\\ 
 \cline{2-14}
 
 &\multicolumn{5}{l}{\cellcolor{yellow}Total For Third Approach } & \cellcolor{green}14.09& 3.35 &\cellcolor{orange}88.81 &0&14.09  & 0 & \cellcolor{orange}1 &\cellcolor{orange}0.27   \\
 \hline
 \multirow{6}{*}{\rotatebox{90} {2nd test Case}} & 1St & 200 &20  &10  &\cellcolor{green}302  &27.44  & 0 & 100 &0 &0 &0  & 0  & 0 \\ 
 \cline{2-14}
 
 &\multicolumn{5}{l}{\cellcolor{yellow}Total For First Approach } & \cellcolor{green}27.44 & 0 & 100 &0& 0 & 0 & 0 &0   \\
 \cline{2-14}

\cline{2-14}
 & 2nd & 200 & 20 & 10 & \cellcolor{green}244 & 24.64 &6.44& 89.81 & 19 & 19.37 & 19.80 &0&0   \\ 
 \cline{2-14}
 &\multicolumn{5}{l}{\cellcolor{yellow}Total For Second Approach } & 24.64 &6.44 &\cellcolor{orange}89.81 &19&19.73& \cellcolor{green}19.80&0  & 0  \\
 \cline{2-14}

\cline{2-14}
 &3rd  & 200 & 20 & 10 &244  & 19.80 &6.22  &91.05 & 0 & 0 &0& 3 &1.24\\ 
 \cline{2-14}
 
 &\multicolumn{5}{l}{\cellcolor{yellow}Total For Third Approach } & \cellcolor{green}19.80& 6.22 & \cellcolor{orange}91.05 &0& 0 & \cellcolor{green}0 &\cellcolor{orange}3  &\cellcolor{orange} 1.24  \\
 \hline

\end{tabular}
 \caption{The metric of arrived applications }
\end{table}

\newpage
\newpage
\subsection{The Algorithm Simulation Code in Github}
Please Find the code of the algorithms' simulation at github repository. 

https://github.com/abo2512d1/ContainerSechulingAlgorithm/tree/main/github.com/
appSchedul/sch\_Projct

\label{appb}

\end{document}